\documentclass[letterpaper,12pt]{article}

\usepackage{dgjournal}          

\usepackage{mathptmx}

\usepackage{graphics}

\usepackage[authoryear,comma,longnamesfirst,sectionbib]{natbib} 

\RequirePackage[OT1]{fontenc}
\RequirePackage{amsthm,amsmath}
\RequirePackage[authoryear]{natbib}
\RequirePackage[colorlinks,citecolor=blue,urlcolor=blue]{hyperref}

\numberwithin{equation}{section}
\theoremstyle{plain}
\newtheorem{theorem}{Theorem}[section]
\newtheorem*{theorem*}{Theorem}
\newtheorem{corollary}[theorem]{Corollary}
\newtheorem*{corollary*}{Corollary}

\newtheorem*{lemma*}{Lemma}

\newtheorem*{remark*}{Remark}

\newtheorem*{algo*}{Algorithm}

\usepackage{hyperref}
\usepackage{pgf,tikz}
\usepackage{lscape}
\usepackage{subfigure}
\usepackage{amssymb,bbold}
\usepackage{amsmath,amsthm, url}
\usepackage{amsfonts}
\usepackage{amssymb}
\usepackage{booktabs}
\usepackage{longtable}
\newcommand\independent{\protect\mathpalette{\protect\independenT}{\perp}}
\def\independenT#1#2{\mathrel{\rlap{$#1#2$}\mkern2mu{#1#2}}}

\usepackage[graphicx]{realboxes}

\begin{document}
\begin{center}
{\Large
	{\sc  Causal mediation analysis in presence of multiple mediators uncausally related}
}
\bigskip

Allan J\'erolon $^{1}$,  Laura Baglietto $^{2}$, \'Etienne Birmel\'e $^{1}$, Vittorio Perduca $^{1}$ \& Flora Alarcon $^{1}$,
\bigskip

{\it
$^{1}$ Laboratory MAP5, Universit\'e Paris Descartes and CNRS,\\ Sorbonne Paris Cit\'e, Paris France, allan.jerolon@parisdescartes.fr, etienne.birmele@parisdescartes.fr, vittorio.perduca@parisdescartes.fr, flora.alarcon@parisdescartes.fr

$^{2}$ Department of Clinical and Experimental Medicine\\
Universit\`a di Pisa, Italy et laura.baglietto@unipi.it
}
\end{center}
\bigskip


{\bf Abstract.} Mediation analysis aims at disentangling the effects of a treatment on an outcome through alternative causal mechanisms and has become a popular practice in biomedical and social science applications. The
causal framework based on counterfactuals is currently the standard approach to mediation, with important methodological advances introduced
in the literature in the last decade, especially for simple mediation, that is
with one mediator at the time. Among a variety of alternative approaches,
K. Imai et al. showed theoretical results and developed an
R package to deal
with simple mediation as well as with multiple mediation involving multiple mediators conditionally independent given the treatment and baseline
covariates. This approach does not allow to consider the often encountered
situation in which an unobserved common cause induces a spurious correlation between the mediators. In this context, which we refer to as mediation with uncausally related mediators, we show that, under appropriate
hypothesis, the natural direct and joint indirect effects are non-parametrically
identifiable. Moreover, we adopt the quasi-Bayesian algorithm developed by Imai et al. and propose a procedure based on the simulation of counterfactual distributions to estimate not only the direct and joint indirect effects but also the indirect effects through individual mediators. We study the properties of the proposed estimators through simulations. As an illustration, we
apply our method on a real data set from a large cohort to assess the effect
of hormone replacement treatment on breast cancer risk through three mediators, namely dense mammographic area, nondense area and body mass index.


\section{Introduction}
\begin{figure}[h]
\begin{center}
\begin{tikzpicture}
\node (t) at (0,0) {$T$};
\node (y) at (5,0) {$Y$};
\node (m) at (2.5,1) {$M$};
 
\draw[->, line width= 1] (t) --  (y);
\draw [->, line width= 1] (m) -- (y);
\draw [->, line width= 1] (t) -- (m);
\end{tikzpicture}
\end{center}
\caption{Simple mediation model with one mediator $M$ and no confounding covariates.}\label{fig0}
\end{figure}
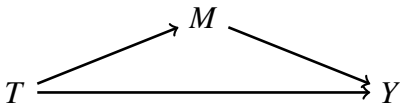

Causal mediation analysis comprises statistical methods to study the mechanisms underlying the relationships between a cause, an outcome and a set of intermediate variables. This approach has become increasingly popular in various domains such as biostatistics, epidemiology and social sciences. Mediation analysis applies to the situation depicted by the causal directed acyclic graph of Figure \ref{fig0}, where an exposure (or treatment) $T$ affects an outcome $Y$ either directly or through one or more intermediate variables referred to as \textit{mediators}. The aim of the analysis is to assess the total causal effect of $T$ on $Y$ by decomposing it into a \textit{direct} effect and an \textit{indirect} effect through the mediator(s).

Mediation analysis originally developed within the setting of linear structural equation modeling (LSEM) (\cite{baron_moderator-mediator_1986, james_causal_1982,mackinnon_introduction_2008}). Following the seminal works by \cite{robins_identifiability_1992} and \cite{pearl_direct_2001}, a formal framework based on counterfactual established itself as the standard approach to mediation analysis, with a growing methodological literature, see for instance (\cite{petersen_estimation_2006,vanderweele2009conceptual,vanderweele_odds_2010,lange2012simple}) and the comprehensive book by \cite{vanderweele2015explanation}.


In this work, we adopt the point of view and formalism of \cite{imai_general_2010,imai_identification_2010}, who put forward a general approach based on counterfactuals to define, identify and estimate causal mediation effects without assuming any specific statistical model in the particular case of a single mediator. Their theoretical results are based on a strong set of assumptions known as \textit{Sequential Ignorability.} These conditions are interpreted as the requirement that there must be no confounding of the $T-Y$, $T-M$ and $M-Y$ relationships after adjustment on the measured pretreatment covariates (i.e. confounder that is not affected by $T$) and $T$, and moreover that there must not be posttreatment confounding (i.e. confounder that is affected by $T$) between $M$ and $Y$ whatsoever, measured or unmeasured. In particular, \cite{imai_identification_2010,imai_general_2010} proved that under Sequential Ignorability, the average indirect effect is non parametrically identified, see Theorem \ref{theo1} in the next section, and proposed a sensitivity analysis to assess the robustness of estimates to violations of Sequential Ignorability. Moreover they introduced estimation algorithms for the effects of interest that are implemented in the widely used \texttt{ mediation } R package (\cite{tingley_mediation_2014}). 

\begin{figure}[htp]
\begin{center}
\minipage{0.4\textwidth}
\subfigure[Independent]{\label{fig1a}

\begin{tikzpicture}
\node (t) at (0,0) {$T$};
\node (y) at (3.5,0) {$Y$};
\node (m) at (1.75,2.5) {$M$};
\node (w) at (1.75,1.1) {$W$};
\node (a) at (1.75,-1.1) {};

\draw[->,line width= 1] (t) --  (y);
\draw [->,line width= 1] (m) -- (y);
\draw [->,line width= 1] (t) -- (m);
\draw [->, line width= 1] (w) -- (y);
\draw [->,line width= 1] (t) -- (w);
\end{tikzpicture}
}
\endminipage
\minipage{0.4\textwidth}
\subfigure[Causally correlated]{\label{fig1b}

\begin{tikzpicture}
 
\node (t2) at (0,0) {$T$};
\node (y2) at (3.5,0) {$Y$};
\node (m2) at (1.75,2.5) {$M$};
\node (w2) at (1.75,1.1) {$W$}; 
\node (b) at (1.75,-1.1) {};

\draw[->,line width= 1] (t2) --  (y2);
\draw [->,line width= 1] (m2) -- (y2);
\draw [->,line width= 1] (t2) -- (m2);
\draw [->, line width= 1] (w2) -- (y2);
\draw [->,line width= 1] (t2) -- (w2);
\draw [->,line width= 1] (m2) -- (w2);

\end{tikzpicture}
}
\endminipage
\minipage{0.4\textwidth}
\subfigure[Uncausally correlated]{\label{fig1c}

\begin{tikzpicture}
\node (t3) at (0,0) {$T$};
\node (y3) at (3.5,0) {$Y$};
\node (m3) at (1.75,2.5) {$M$};
\node (w3) at (1.75,1.1) {$W$};
\node (c.) at (1.75,-1.1) {};



\draw[->,line width= 1] (t3) --  (y3);
\draw [->,line width= 1] (m3) -- (y3);
\draw [->,line width= 1] (t3) -- (m3);
\draw [->, line width= 1] (w3) -- (y3);
\draw [->,line width= 1] (t3) -- (w3);
\draw [<->,dashed,line width= 1] (m3) -- (w3);

\end{tikzpicture}
}
\endminipage

\caption{Three situations with multiple mediators $M$ and $W$.}\label{multiplecases}
\end{center}
\end{figure}
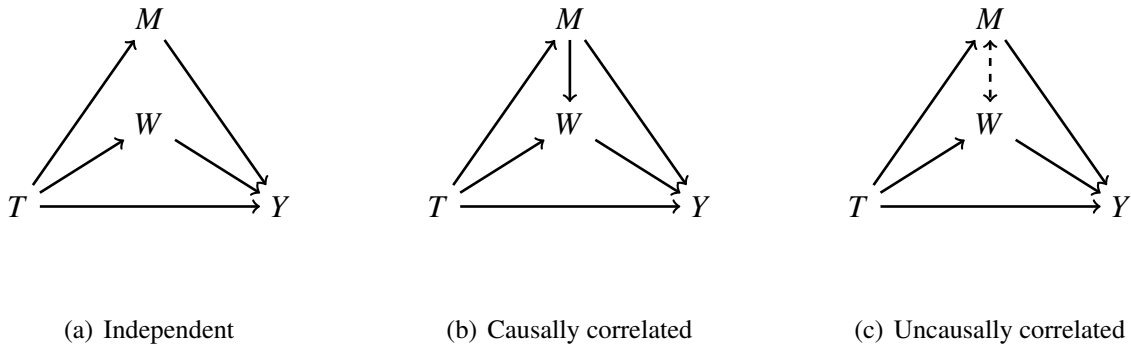

When multiple mediators are involved in the mediation model, three cases may arise, as shown in Figure \ref{multiplecases}: in Fig. \ref{fig1a} mediators are conditionally independent given the treatment and measured covariates (not depicted here), in Fig. \ref{fig1b} mediators are causally ordered, that is one affects the other; in Fig. \ref{fig1c} mediators are conditionally dependent given the treatment and measured covariates without being causally ordered. In the latter situation, we will talk about \textit{uncausally correlated} mediators as opposed to the situation of Fig. \ref{fig1b} where mediators are causally correlated. We will also refer to the cases depicted in figures \ref{fig1a} and \ref{fig1c} as mediation with multiple \textit{causally unrelated} mediators.

Models in Figures \ref{fig1a} and \ref{fig1b} have been treated in the last few years (\cite{vanderweele_mediation_2014,lange_assessing_2014,daniel_causal_2015}) and will be commented further in the discussion section.

\begin{figure}[htp]
\begin{center}

\begin{tikzpicture}
\node (t) at (0,0) {$T$};
\node (y) at (3.5,0) {$Y$};
\node (m) at (1.75,2.5) {$M$};
\node (w) at (1.75,1.1) {$W$};
\node (u) at (3.5,2.5) {$U$};

\draw[->,line width= 0.5] (u) --  (m);
\draw[->,line width= 0.5] (u) --  (w);
\draw[->,line width= 1] (t) --  (y);
\draw [->,line width= 1] (m) -- (y);
\draw [->,line width= 1] (t) -- (m);
\draw [->, line width= 1] (w) -- (y);
\draw [->,line width= 1] (t) -- (w);
\end{tikzpicture}
\end{center}
\caption{Correlation between mediators due to $U$.}\label{fig3}
\end{figure}
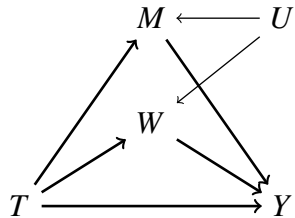

  Figure \ref{fig1c} corresponds to an Acyclic Directed Mixed Graph (ADMG) as introduced by \cite{shpitser_counterfactual_2013} and \cite{shpitser_indentification_2018}. Bidirected dotted edges indicate a non-causal correlation, due for instance to a latent common cause, as in Figure~\ref{fig3}. Shpitser and coauthors define districts as the connected components of the graph restricted to the bidirected edges and describe a necessary and sufficient condition for effects to be identified, that is expressed in terms of observational data. In the case of a multiple mediation framework, this condition says that the effect mediated by a set $\mathcal{S}$ of mediators can be written as a function of the observations if and only if $\mathcal{S}$ can be written as the union of some districts. In the case of Figure~\ref{fig1c}, it means that the direct effect (mediated by neither $M$ nor $W$) and the joint effect (mediated by both $M$ and $W$) can be written in terms of observations, but that the effect mediated only by $M$ cannot.

 The estimation of such individual indirect effects, each specific to a given mediator, is however of practical importance. To do so, \cite{imai_identification_2013} extend their above mentioned approach to multiple mediators. When mediators are causally unrelated,
 and Sequential Ignorability holds, they suggested to process several single mediator analyses in parallel, one mediator at the time. Obviously, this approach leads to a biased estimate of the direct effect, because it forces the indirect effects via all other mediators to contribute to the direct effect. More subtly, this approach is not appropriate when mediators are uncausally correlated due to an unmeasured covariate $U$ causally affecting both mediators $M$ and $W$ as in Figure \ref{fig3}. As a matter of fact, in this situation $U$ is an unobserved confounder of the relationship between $M$ and $Y$ and Sequential Ignorability does not hold. This key fact was remarked by \cite{imai_identification_2013} and \cite{vanderweele_mediation_2014}, but no explicit solution to the problem was proposed other then conducting the above mentioned sensitivity analysis.
 
In this article, we focus on the scenario of multiple causally unrelated mediators (either independent, Fig. \ref{fig1a}, or uncausally correlated, Fig. \ref{fig1c}). First, we extend the theoretical results developed by Imai and coauthors to this scenario.  To do so, we show that, under assumptions alternative to Sequential Ignorability, the direct effect and the joint indirect effect through the vector of all mediators can be expressed by a formula involving observed variables only, while the indirect effect through each individual mediator is given by a formula involving both observed and counterfactual variables. The first formulas lead to an unbiased estimation of the direct and joint indirect effects, compliantly to \cite{shpitser_indentification_2018}. In addition, we propose a procedure based on the simulation of counterfactual distributions to estimate the indirect effects through individual mediators. As the proposed estimation method depends on the simulation of counterfactuals and not only on the observational data, we  conduct an empirical study to show that the method result in unbiased estimates of the direct and indirect effects.  Our methods is implemented in R (a documented R package is under preparation and will be soon posted on GitHub). Finally, we apply our method to a real dataset from a large cohort to assess the effect of hormone replacement treatment on breast cancer risk through three uncausally correlated mediators, namely dense mammographic area, nondense area and body mass index.

\medskip

 For sake of clarity, we list the notations used in this article:

\begin{itemize}
\item $T \in \{0,1\}$ : treatment 
\item $Z \in \mathbb{R}^K$: vector of all mediators
\item $M^k \in \mathbb{R}$:  $k$-th mediator, when this is clear from the context we will use the notation $M=M^k$
\item $W^k \in \mathbb{R}^{K-1}$ : complement of $M^k$ in $Z$, when this is clear from the context we will use the notation $W=W^k$
\item $X \in \mathbb{R}^P$: vector of pretreatment confounders 
\item $Y \in \mathbb{R}$ or $\{0,1\}$ : outcome
\item $\delta^k(t)$: indirect effect of $T$ mediated by $M^k$
\item $\delta(t)$: indirect effect of $T$ mediated by $M$
\item $\zeta(t)$: direct effect of $T$ 
\item $\delta, \zeta$: averages $(\delta(0)+\delta(1))/2$ and $(\zeta(0)+\zeta(1))/2$
\item $\tau$: total effect
\item $PM^k(t)=\delta^k(t)/\tau$: proportion mediated by $M^k$ 
\item $\Phi$: the cumulative distribution function of the standard normal distribution $\mathcal{N}(0,1)$
\item $A^{\Gamma}$: the transpose of a matrix or vector $A$.
\end{itemize} 


\section{Brief review of simple mediation}

\begin{figure}[h]
\begin{center}
\begin{tikzpicture}
\node (t) at (0,0) {$T$};
\node (y) at (5,0) {$Y$};
\node (m) at (2.5,1) {$M$};
\node (x) at (0,1.5) {$X$};
 
\draw[->, line width= 1] (t) --  (y);
\draw [->, line width= 1] (m) -- (y);
\draw [->, line width= 1] (t) -- (m);
\draw [->, line width = 0.5] (x) -- (t);
\draw [->, line width = 0.5] (x) -- (m);
\draw [->, line width = 0.5] (x) to [bend left=45] (y);
\end{tikzpicture}
\end{center}
\caption{Simple mediation causal diagram. }\label{fig1}
\end{figure}
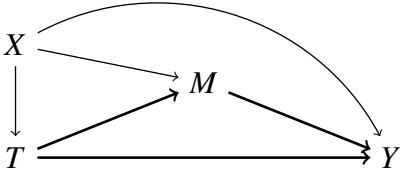

We begin by recalling the main results by \cite{imai_identification_2010} in the case of a simple mediator and a binary treatment; we will adopt the same notations. Let $Y$ be the variable denoting the observed outcome, $T$ the treatment or exposure (coded as 1 for treated or exposed and 0 for non-treated or non-exposed) and $M$ a single intermediate variable on the causal path from the $T$ to $Y$. Finally let $X$ represent a vector of pretreatment confounders. The causal diagram in Figure \ref{fig1} depicts the causal relation between the four variables.

\bigskip

 The causal approach to mediation analysis requires two types of counterfactual variables. On one hand, we consider the potential mediator when the treatment is set to $t$, denoted $M(t)$. On the other hand,  we consider the potential outcome when the treatment is set to $t$ and the mediator has the potential value when the treatment is set to $t'$, denoted $Y(t,M(t'))$. We recall the definition of counterfactuals in  the \href{http://helios.mi.parisdescartes.fr/~ajerolon/indexang.html}{supplementary materials}.

 The three quantities of interest in simple mediation analysis are the average causal indirect effect denoted $\delta(t)$, the average direct effect $\zeta(t)$, for $t \in \{0,1\}$, and the average total effect $\tau$: 

\begin{eqnarray*}
\delta(t) & = & \mathbb{E}[Y(t,M(1))|X] - \mathbb{E}[Y(t,M(0))|X]\label{d1}\\
\zeta(t) & = & \mathbb{E}[Y(1,M(t))|X] - \mathbb{E}[Y(0,M(t))|X]\label{z1}\\
\tau & = & \mathbb{E}[Y(1,M(1))|X] - \mathbb{E}[Y(0,M(0))|X].\label{t1}
\end{eqnarray*}


 Imai and collaborators showed in the theorem below that these effects can be identified regardless of a model assumption under two crucial hypothesis that go under the name of Sequential Ignorability Assumption (SIA):  

\begin{eqnarray}
\{Y(t',m),M(t)\} \independent T|X=x \quad \forall\, t,t',m\label{ass1}\\
Y(t',m) \independent M(t)|T,X=x \quad \forall\, t,t',m. \label{ass2}
\end{eqnarray}

\begin{theorem}[\cite{imai_identification_2010}]\label{theo1}
Under SIA, the average indirect effect and the direct effect are identified non-parametrically and are given by, for $ t\in\{0,1\}$,
\begin{eqnarray*}
\delta(t) & = & \int \int \mathbb{E}[Y|M=m,T=t,X=x] \mathrm{d}F_{M|T=1,X=x}(m)\\
&  & -\int \mathbb{E}[Y|M=m,T=t,X=x]\mathrm{d}F_{M|T=0,X=x}(m)\mathrm{d}F_{X}(x)\\
\zeta(t) & = & \int \int \mathbb{E}[Y|M=m,T=1,X=x] \mathrm{d}F_{M|T=t,X=x}(m)\\
&  & -\int \mathbb{E}[Y|M=m,T=0,X=x] \mathrm{d}F_{M|T=t,X=x}(m)\mathrm{d}F_{X}(x).
\end{eqnarray*}
\end{theorem}

 In the setting of linear models, the two corollaries below follow, the first for a continuous outcome and the second for a binary outcome.

\begin{corollary}[\cite{imai_identification_2010}]\label{cor1}
 Under SIA and assuming the following linear structural equation model (LSEM)

\begin{eqnarray*}
M & = & \alpha_2 + \beta_{2}T + \xi^{\Gamma}_2 X + \epsilon_2\\
Y & = & \alpha_{3}+\beta_{3}T + \gamma M + \xi^{\Gamma}_3 X + \epsilon_3,
\end{eqnarray*}
where $\epsilon_{i}\sim\mathcal{N}(0,\sigma_i^2)$ for $i\in\{2,3\}$, the average indirect and direct effects are identified by $\delta(0)=\delta(1)=\beta_2\gamma$ and $\zeta(0)=\zeta(1)=\beta_3$.
\end{corollary}

\medskip

 In the situation of a binary outcome, two main alternatives exist to model its  conditional distribution. On the one hand we can consider the probit regression

$$
\mathbb{P}(Y=1|T,M,X) = \Phi_{\mathcal{N}(0,\sigma_3^2)}(\alpha_{3}+\beta_{3}T +\gamma M+\xi_3^{\Gamma}X), 
$$
 where $\Phi_{\mathcal{N}(0,\sigma_3^2)}$ is the cumulative density function of normal distribution $\mathcal{N}(0,\sigma_3^2)$.

 On the other hand we can assume the logistic regression $$\text{logit } (\mathbb{P}(Y=1|T,M,X)) = \alpha_{3}+\beta_{3}T +\gamma M +\xi_3^{\Gamma}X. $$

\begin{corollary}[\cite{imai_identification_2010}]\label{cor2}
Let $Y$ be binary and assume the model
\begin{eqnarray*}
M & = & \alpha_2 + \beta_2 T + \xi_2^{\Gamma } X + \epsilon_2 \nonumber\\
Y & = & \mathbb{1}_{ \{Y^*>0\} } \mbox{, where } Y^{*} =  \alpha_{3}+\beta_{3}T + \gamma M +\xi_3^{\Gamma}X + \epsilon_3 \\
\end{eqnarray*}
 where $\epsilon_{2}\sim\mathcal{N}(0,\sigma_2^2)$ and $\epsilon_3\sim\mathcal{N}(0,\sigma_3^2)$ (probit regression) or $\epsilon_3\sim\mathcal{L}(0,1)$ (logit regression), where $\mathcal{L}(0,1)$ denotes the standard logistic distribution.

 Under SIA, the average indirect and direct effects are identified by 
\begin{eqnarray*}
\delta(t) & = & \mathbb{E}[F_u(h_{t,1})-F_u(h_{t,0})|X] \\ 
\zeta(t) & = & \mathbb{E}[F_u(h_{1,t})-F_u(h_{0,t})|X] 
\end{eqnarray*}
where 
$$
h_{t,t'} = \alpha_{3}+\beta_{3}t +\gamma (\alpha_2 + \beta_{2} \times t'+ \xi_2^{\Gamma} X)+\xi_3^{\Gamma} X
$$ 
and for a probit regression the function $F_u$ is 
$$
F_u(z)= \Phi\left(\dfrac{z}{\sqrt{\gamma^2\sigma_2^2+1}}\right)
$$
while for a logit regression we have
$$
F_u(z)= \int^{\infty}_{-\infty} \Phi\left(\dfrac{z-y}{\gamma\sigma_2}\right) \dfrac{e^{y}}{(1+e^{y})^2} \, \mathrm{d}y.
$$
\end{corollary}

\section{Extension to multiple causally unrelated mediators}
\label{section:mcum}

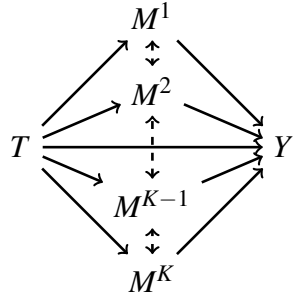
\begin{figure}[htp]
 \begin{center}
 \begin{tikzpicture}
\node (t) at (0,0) {$T$};
\node (y) at (3.5,0) {$Y$};
\node (m1) at (1.75,1.75) {$M^1$};
\node (m2) at (1.75,0.75) {$M^2$};
\node (mK1) at (1.75,-0.75) {$M^{K-1}$};
\node (mK) at (1.75,-1.75) {$M^K$};
 
\draw[->, line width= 1] (t) --  (y);
\draw [->, line width= 1] (m1) -- (y);
\draw [->, line width= 1] (t) -- (m1);
\draw [->, line width= 1] (m2) -- (y);
\draw [->, line width= 1] (t) -- (m2);
\draw [->, line width= 1] (mK) -- (y);
\draw [->, line width= 1] (t) -- (mK);
\draw [->, line width= 1] (mK1) -- (y);
\draw [->, line width= 1] (t) -- (mK1);
\draw [<->, dashed, line width =1] (m1) -- (m2);
\draw [<->, dashed, line width =1] (mK1) -- (m2);
\draw [<->, dashed, line width =1] (mK) -- (mK1);

\end{tikzpicture}
\caption{Multiple mediation causal diagram with correlated mediators. The vector of pretreatment confounders $X$ is not shown. Dashed lines represent possible non-causal correlations and solid lines causal relationships. Uncausally correlations is possible between each pair of mediators but for the readability of the figure, they do not appear.}\label{figmul}
\end{center}
\end{figure}

In this subsection, we consider that  $K$ mediators intervene in the causal relationship between $T$ and $Y$  as in Figure \ref{figmul}. In particular, the following definitions and results apply when mediators are independent (Figure \ref{fig1a}) or uncausally correlated (Figure \ref{fig1c}).
\medskip

\subsection{Effect definitions.}\label{definitionmultiple} 

\medskip

 Let $Z$ be the vector of all $K\geq 2$ mediators and $M^k$ the mediator of interest. We denote by $W^k$ the complement of $M^k$ in $Z$, that is all mediators that are not of direct interest, and $X$ the vector of pretreatment confounders. An illustration is given in Figure \ref{figmul}. 

 The average indirect effect mediated by $M^k$ was defined by  \cite{imai_identification_2013} as 
\begin{eqnarray*}
\delta^k(t) =  \mathbb{E}\left[Y(t,M^k(1),W^k(t))|X\right] - \mathbb{E}\left[Y(t,M^k(0),W^k(t))|X\right]. \label{dk}
\end{eqnarray*}

 As a measure of the average joint indirect effect, that is the indirect effect mediated by all the mediators, we take
\begin{eqnarray*}
\delta^Z(t) =  \mathbb{E}\left[Y(t,Z(1))|X\right] - \mathbb{E}\left[Y(t,Z(0))|X\right].\label{dz}
\end{eqnarray*}

\begin{remark*}
 Note that the joint indirect effect can be decomposed as
$$
\delta^Z(t) = \dfrac{\sum_{k=1}^K \left(\delta^k(t)+\eta^k(t)\right) }{K}
$$
 where
$$
\eta^k(t) =  \mathbb{E}\left[Y(t,M^k(1-t),W^k(1))|X\right]-\mathbb{E}\left[Y(t,M^k(1-t),W^k(0))|X\right].
$$
A proof for this result can be found in the Appendix \ref{link}.
\end{remark*}


 Each of the $2^K$ direct effects is defined as 
\begin{eqnarray*}
\zeta(t_1,\dots,t_K) & = & \mathbb{E}\left[Y(1,M^1(t_1),\dots,M^K(t_K))|X\right] \\
&  & -\mathbb{E}\left[Y(0,M^1(t_1),\dots,M^K(t_K))|X\right] \label{zk}
\end{eqnarray*}

 where $t_k \,\in \, \{0,1\}$ for all $k \, \in \, \{1,\dots,K\}$.\\

 For the sake of simplicity, among all these direct effects we will consider only $\zeta(0,\dots,0)$ and $\zeta(1,\dots,1)$, denoted $\zeta(t), \, t\in\{0,1\}$.

 The total effect $\tau$ is
\begin{eqnarray*}
\tau & = & \mathbb{E}\left[Y(1,Z(1))|X\right] - \mathbb{E}\left[Y(0,Z(0))|X\right].\label{tk}
\end{eqnarray*}

  Note that $\tau$ is the sum of the joint indirect effect of treatment $t$ and of the direct effect of treatment $1-t$: 
\begin{eqnarray*}
\tau & = & \delta^Z(t) + \zeta(1-t).
\end{eqnarray*}

\subsection{Assumptions}\label{assumptions}
 Our results are based on the following hypothesis that we called Sequential Ignorability for Multiple Mediators Assumption (SIMMA):

$$\begin{array}{rrll}
\eqref{ISB1} \quad & \{Y(t,m,w),M(t'),W(t'')\} & \independent & T|X=x  \\
\eqref{ISB4} \quad & Y(t,m,w) & \independent & \left(M(t'),W(t'')\right)|T,X=x
\end{array}$$

for all possible values of $t,t',t'',m,w$. A detailed explanation of SIMMA can be found in Appendix \ref{hypothese}.

\medskip

Here we recall that $X$ is the vector of all observed pretreatment covariates, that is variables that are not affected by the treatment. The first hypothesis implies that there must not be any unobserved pretreatment confounders between the treatment and the outcome and between the treatment and the individual mediators once conditioning on all observed covariates. The second hypothesis excludes the existence of two distinct types of confounding between the mediators taken \textit{jointly} and the outcome: the confounding by an unobserved pretreatment variable and the confounding by an observed or unobserved postreatment variable. 

Crucially, this second hypothesis replaces the second and third hypothesis of \cite{imai_identification_2013} in the situation of multiple causally independent mediators, where this requirement applies to each mediator separately (see Appendix \ref{hypothese} for more details about the comparison between the two sets of hypothesis). 

Note that this is not the only important difference with Imai's assumptions. As a matter of fact, in this article we are interested in the situation where $M$ and $W$ are uncausally correlated, typically because of a pretreatment variable $U$ affecting both as in Figure \ref{fig5a}. Note that if $U$ is unobserved (i.e. it is not taken into account by $X$) the second equation of SIMMA is not violated because the joint distribution of the mediators incorporates the influence of $U$ on the individual mediators. On the contrary, such a $U$ would violate the corresponding hypothesis by \cite{imai_identification_2013} because it constitues an unobserved confounder of the relations between $W$ and $Y$ and $M$ and $Y$.

\medskip

\subsection{Identifiability} \label{ssection:ident}
 Note that the mediator of interest $M$ can be any of the $K$ mediators, so that the following results can be applied to each mediator. In particular, this will allow to provide the indirect effect by each mediator taken individually. 

\medskip

 Our first result extends Theorem \ref{theo1} to multiple mediators, not only when mediators are causally independent as done by \cite{imai_identification_2013}, but also when they are uncausally correlated.

\begin{theorem}\label{theo1J} Under SIMMA and assuming $K$ mediators that can be either independent or uncausally correlated, the following results hold:

\medskip

 The average indirect effect of the mediator of interest is given by:

\begin{equation}
\begin{array}{ccl}
  \delta(t) & = & \displaystyle\int \int_{\mathbb{R}^K} \mathbb{E}\left[Y|M=m,W=w,T=t,X=x\right] \\
  &  & \{\mathrm{d}F_{\left(M(1),W(t)\right)|X=x}(m,w)-\mathrm{d}F_{\left(M(0),W(t)\right)|X=x}(m,w)\} \mathrm{d}F_{X}(x).  
\end{array} \label{eq:delta_M}
 \end{equation}

\medskip

 Moreover the joint indirect effect, the direct effect and the total effect are respectively identified non-parametrically by:
 
 \begin{eqnarray*}
  \delta^Z(t) & = & \int \int_{\mathbb{R}^K} \mathbb{E}\left[Y|Z=z,T=t,X=x\right] \mathrm{d}F_{Z|T=1,X=x}(z) \\
  &  & - \int_{\mathbb{R}^K} \mathbb{E}\left[Y|Z=z,T=t,X=x\right]\mathrm{d}F_{Z|T=0,X=x}(z) \mathrm{d}F_{X}(x),
  \end{eqnarray*}
  
\begin{eqnarray*}
 \zeta(t) & = & \int \int_{\mathbb{R}^K} \mathbb{E}(Y|Z=z,T=1,X=x) \mathrm{d}F_{Z|T=t,X=x}(z)\\
 &  & - \int_{\mathbb{R}^K}\mathbb{E}(Y|Z=z,T=0,X=x) \mathrm{d}F_{Z|T=t,X=x}(z) \mathrm{d}F_{X}(x), \nonumber
\end{eqnarray*}

\begin{eqnarray*}
 \tau & = & \int \left(  \int_{\mathbb{R}^K} \mathbb{E}(Y|Z=z,T=1,X=x) \mathrm{d}F_{Z|T=1,X=x}(z) \right. \\
 &  & \left. -\int_{\mathbb{R}^K} \mathbb{E}(Y|Z=z,T=0,X=x) \mathrm{d}F_{Z|T=0,X=x}(z)\right) \mathrm{d}F_{X}(x). \nonumber
\end{eqnarray*}
\end{theorem}

\medskip

 Theorem \ref{theo1J} has the same role in multiple mediation as Theorem \ref{theo1} in simple mediation, because it shows that under proper assumptions, the (joint) indirect and direct effects are nonparametrically identified. In particular, note that the last two equations make it possible to derive estimators for the joint indirect effect and for the direct effect, as already shown by \cite{shpitser_counterfactual_2013}. However, equation (\ref{eq:delta_M}) does not allow to derive an estimator of the individual indirect effect of the mediator of interest, because the conditional distribution of $\left(M(t'),W(t)\right)$ is not observable. Note that in the particular case where $M$ is independent of $W$, equation (\ref{eq:delta_M}) becomes
\begin{eqnarray*}
  \delta(t) & = & \int \int \mathbb{E}\left[Y|M=m,T=t,X=x\right] \mathrm{d}F_{M|T=1,X=x}(m)\\
  &  & - \int \mathbb{E}\left[Y|M=m,T=t,X=x\right]\mathrm{d}F_{M|T=0,X=x}(m) \mathrm{d}F_{X}(x),
\end{eqnarray*}
which the same equation for $\delta(t)$ given by Theorem \ref{theo1}, thus allowing to identify the average indirect effect non-parametrically. This result was reported by \cite{imai_identification_2013}. A proof of Theorem \ref{theo1J} can be found in appendix \ref{ptheo1J}.

\bigskip

 The following two corollaries show identification formulas for the indirect and direct effects in the setting of the LSEM or when the mediating variables are gaussian and $Y$ is binary. 

\medskip

 Crucially, in the following corollaries we assume that correlations between the potential mediators are the same whatever the treatment governing the mediators:

$$cor\left(M^i(t),M^j(t')|T,X\right)= \rho_{ij}, \forall \, t,t' \in \{0,1\}, \, \forall \, i,j \in [1,K].$$

\subsection{Continuous outcome}
\begin{corollary}\label{cor1J}
With $K$ mediators and $P$ covariables we assume the following linear model
\begin{eqnarray}
Z & = & \alpha_2 + \beta_2 T + \xi_2^{\Gamma} X + \Upsilon_2 \label{linMk}  \\
Y & = & \alpha_{3}+\beta_{3}T + \gamma^{\Gamma} Z +\xi_3^{\Gamma}X + \epsilon_3, \label{linY} 
\end{eqnarray}
where $\alpha_2,\beta_2,\gamma\in\mathbb{R}^K$, $\xi_2^{\Gamma} \in\mathbb{R}^K\times\mathbb{R}^P$, $\xi_3^\Gamma\in\mathbb{R}^P$,\,\text{and} \, $\Upsilon_2\sim\mathcal{N}(0,\Sigma_2)$ is the vector of residuals with covariance matrix $\Sigma_2\in\mathbb{R}^K\times\mathbb{R}^K$ and $\epsilon_3\sim\mathcal{N}(0,\sigma^2_3)$, with $\sigma_3\in\mathbb{R}$. 

 We assume that the K mediators are either independent or non-causally correlated. In the latter case, we assume that pairwise correlations between potential mediators do not depend on the treatments governing them. Under SIMMA the indirect effect of the $k$-th mediator is identified and given by:
$$
\delta^k(0)=\delta^k(1)=\gamma_k\beta^k_2.
$$ 

 Moreover, the joint indirect effect is the sum of the average indirect effects by each mediator:

$$
\delta^Z(t)  = \sum_{k=1}^K \delta^k(t). 
$$

 The direct effect of the $k$-th mediator is also identified and given by
$$
\zeta(0)=\zeta(1)=\beta_3.
$$
\end{corollary}

\medskip


 A proof of Corollary \ref{cor1J} can be found in the \href{http://helios.mi.parisdescartes.fr/~ajerolon/indexang.html}{supplementary materials}. Note that an equivalent result for the joint indirect effect is shown in \cite{vanderweele_mediation_2014}.

\medskip

 We have already observed that if the $K$ mediators are independent, the equation for the marginal indirect effect given by Theorem \ref{theo1J} (multiple analysis) reduces to the equation given by Theorem \ref{theo1} (simple analysis). In this situation, Corollary \ref{cor1J} implies that in the LSEM setting, the  indirect effects given by simple analyses can be summed up to obtain the joint indirect effect. Obviously, simple analyses do not allow to assess a comprehensive direct effect, because depending on the mediator of interest, each simple analysis will lead to a different direct effect. All these aspects will be illustrated through simulations in Section \ref{simulsection}.  

\medskip
\subsection{Binary outcome}
 We now address the case of a binary outcome. As for simple mediation, we consider either the probit regression

$$
\mathbb{P}(Y=1|T,Z,X) = \Phi_{\mathcal{N}(0,\sigma_3^2)}(\alpha_{3}+\beta_{3}T +\gamma^{\Gamma} Z +\xi_3^{\Gamma}X), 
$$

 or the logistic regression
$$
\text{logit } (\mathbb{P}(Y=1|T,Z,X)) = \alpha_{3}+\beta_{3}T +\gamma^{\Gamma} Z +\xi_3^{\Gamma}X. 
$$
\begin{corollary}\label{cor2J}

Assume the following model with a binary outcome :
\begin{eqnarray}
Z  & = & \alpha_2 + \beta^{\Gamma}_2 T + \xi_2^{\Gamma} X + \Upsilon_2, \,  \label{binM2}\\
Y^{*}  & =  &\alpha_{3}+\beta_{3}T +\gamma^{\Gamma} Z +\xi_3^{\Gamma}X + \epsilon_3, \,  \label{binM3}\\
Y  & = & \mathbb{1}_{ \{Y^*>0\} }
\end{eqnarray} 
 
  where $\Upsilon_2\sim\mathcal{N}(0,\Sigma_2)$ and where
 $\epsilon_3\sim\mathcal{N}(0,\sigma_3^2)$ or  $\mathcal{L}(0,1)$. We assume that the K mediators are either independent or non-causally correlated.  In the latter case, we assume that pairwise correlations between potential mediators do not depend on the treatments governing them. Under SIMMA, the effects of interest are given by:

\begin{eqnarray}
\delta^k(t) & = & \int F_U\left((\alpha_3 + \sum_{j=1}^K\gamma_j\alpha^j_2) + (\beta_3 + \sum_{j=1,j\neq k}^{K}\gamma_j\beta^j_2)t + \gamma_k\beta^k_2 \times 1 + (\xi_3 + \sum_{j=1}^K\gamma_j\xi_2^{\Gamma j})x\right) \nonumber\\
     &  & - F_U\left((\alpha_3 + \sum_{j=1}^K\gamma_j\alpha^j_2) + (\beta_3 + \sum_{j=1,j\neq k}^{K}\gamma_j\beta^j_2)t + \gamma_k\beta^k_2 \times 0 + (\xi_3 + \sum_{j=1}^K\gamma_j\xi_2^{\Gamma j})x\right) \mathrm{d}X, \nonumber \\
\delta^Z(t) & = & \int F_U\left((\alpha_3 + \sum_{k=1}^K\gamma_k\alpha^k_2) + \beta_3 \times 1 + \sum_{k=1}^K\gamma_k\beta^k_2t + (\xi_3 + \sum_{k=1}^K\gamma_k\xi_2^{\Gamma k})x\right) \nonumber\\
     &  & - F_U\left((\alpha_3 + \sum_{k=1}^K\gamma_k\alpha^k_2) + \beta_3 \times 0 + \sum_{k=1}^K\gamma_k\beta^k_2t  + (\xi_3 + \sum_{k=1}^K\gamma_k\xi_2^{\Gamma k})x\right) \mathrm{d}X, \nonumber \\
\zeta(t) & = & \int F_U\left((\alpha_3 + \sum_{k=1}^K\gamma_k\alpha^k_2) + \beta_3 \times 1 + (\sum_{k=1}^K\gamma_k\beta^k_2) \times t + (\xi_3 + \sum_{k=1}^K\gamma_k\xi_2^{\Gamma k})x\right) \nonumber \\
& & -F_U\left((\alpha_3 + \sum_{k=1}^K\gamma_k\alpha^k_2) + \beta_3 \times 0 + (\sum_{k=1}^K\gamma_k\beta^k_2) \times t + (\xi_3 + \sum_{k=1}^K\gamma_k\xi_2^{\Gamma k})x\right) \mathrm{d}X, \nonumber 
\end{eqnarray}

 where for a probit regression we have
\begin{equation*}
F_U(z) =  \Phi\left(\dfrac{z}{\sqrt{\sigma_3^2+\displaystyle\sum_{k=1}^K\sum_{j=1}^K\gamma_k\gamma_j cov(\epsilon_2^k,\epsilon_2^j)}}\right), \label{fu1}
\end{equation*}

 and for a logit regression we have
\begin{equation*}
F_U(z)  =   \int_{\mathbb{R}} \Phi\left(\dfrac{z-e_3}{\sqrt{\displaystyle\sum_{k=1}^K\sum_{j=1}^K\gamma_k\gamma_j cov(\epsilon_2^k,\epsilon_2^j)}}\right) \dfrac{e^{e_3}}{(1+e^{e_3})^2}\, \mathrm{d}e_3.\label{fu2}
\end{equation*}

\end{corollary}

 When the mediators are independent we have for a probit regression
\begin{equation*}
F_U(z)  =  \Phi\left(\dfrac{z}{\sqrt{\sigma_3^2+\displaystyle\sum_{k=1}^K\gamma_k^2\sigma_2^2}}\right)\label{fu3},
\end{equation*}
and for a logistic regression
\begin{equation*}
F_U(z)  =   \int_{\mathbb{R}} \Phi\left(\dfrac{z-e_3}{\sqrt{\displaystyle\sum_{k=1}^K\gamma_k^2\sigma_2^2}}\right) \dfrac{e^{e_3}}{(1+e^{e_3})^2}\, \mathrm{d}e_3.\label{fu4}
\end{equation*}

A proof of Corollary \ref{cor2J} can be found in \href{http://helios.mi.parisdescartes.fr/~ajerolon/indexang.html}{supplementary materials}.


\subsection{Estimation algorithm}
\label{ssec:algo}
From the results of section 3.\ref{ssection:ident} we derive estimators of the effects of interest for different kinds of variable. We adapt the quasi-Bayesian algorithm presented by \cite{imai_general_2010},  to the situation of multiple mediators uncausally related, i.e. for independent and uncausally correlated mediators. 

\begin{algo*} Estimate effects of interest:

\begin{enumerate}
\item Fit parametric models for the observed outcome (given all the mediators, treatment and covariates), and mediators (given all the treatment and covariates), denoted respectively as $\widehat{\Theta}_Z= \left(\widehat{\Theta}^1,\dots,\widehat{\Theta}^K\right)$ and $\widehat{\Theta}_Y$.

\item For each model, sample $N$ values for each of its parameters according to their multivariate sampling distribution, denoted as $\widehat{\Theta}_{Z(n)}= \left(\widehat{\Theta}^1_{(n)},\dots,\widehat{\Theta}^K_{(n)}\right)$ and $\widehat{\Theta}_{Y(n)}$, $n=1,\dots,N$. As in (\cite{imai_general_2010}) we use the approximation based on the multivariate normal distribution, with mean and variance equal to the estimated parameters and their estimated asymptotic covariance matrix, respectively.

\item For each $r=1,\dots,R$, repeat the followings steps:

\begin{itemize}
\item Simulate the potential values of each mediator. In particular, for each pair $t,t' \in\{0,1\}$, sample I values, denoted as $Z^k_{(ri)}(t,t')=\left(M^k_{(ri)}(t),W^k_{(ri)}(t')\right)$. When all mediators have the same treatment value, the vector of mediator will be denoted as $Z_{(ri)}(t)=\left(M^k_{(ri)}(t),W^k_{(ri)}(t)\right)$. Note that it is at this step that we take into account the correlation between mediators. 
\item Simulate the potential outcomes given the simulated values of the potential mediators, denoted as $Y_{(ri)}\left(t,Z_{(ri)}^{k}(t',t)\right)$ for each $k$ and $t,t' \in \{0,1\}$.

\item Estimate the causal mediation effects:

\begin{eqnarray*}
\hat{\delta}^{k}_{(r)}(t) & = &  \dfrac{1}{I} \sum^I_{i=1} \left\{Y_{(ri)}\left(t,Z^k_{(ri)}(1,t)\right)- Y_{(ri)}\left(t,Z^k_{(ri)}(0,t)\right)\right\}\\ 
\hat{\delta}^{Z}_{(r)}(t) & = & \dfrac{1}{I} \sum^I_{i=1} \left\{Y_{(ri)}\left(t,Z_{(ri)}(1)\right)- Y_{(ri)}\left(t,Z_{(ri)}(0)\right)\right\}\\
\hat{\zeta}_{(r)}(t) & = & \dfrac{1}{I} \sum^I_{i=1} \left\{Y_{(ri)} \left(1,Z_{(ri)}(t)\right)- Y_{(ri)}\left(0,Z_{(ri)}(t)\right)\right\}\\
\hat{\tau}_{(r)}(t) & = & \dfrac{1}{I} \sum^I_{i=1} \left\{Y_{(ri)} \left(1,Z_{(ri)}(1)\right)- Y_{(ri)}\left(0,Z_{(ri)}(0)\right)\right\}.
\end{eqnarray*} 

\end{itemize}
\item From the empirical distribution of each effect above, obtain point estimates together with p-values and confidence intervals.
\end{enumerate}
\end{algo*}

 Note that this algorithm does not implement the formulas given for the specifc models of Corollaries \ref{cor1J} and \ref{cor2J}.

\medskip

 We have implemented a R  function \texttt{mutimediate()} based on this algorithm and on the function \texttt{mediate()} of the package \texttt{mediation} (\cite{tingley_mediation_2014}). As said in the introduction, a documented R package is under preparation and will be soon posted on GitHub.

\section{Simulation studies}\label{simulsection}

In this section we validate our methodological results through empirical studies. In particular, we compare our estimates of the mediation causal effects to the true effects and to the estimates obtained by running simple mediation analyses, one for each mediator. 

\medskip

\subsection{Data simulation method}\label{simmethod}
 Except for the LSEM framework, it is in general not straightforward to obtain true values of the mediation effects from a causal generative model, that is a set of causal structural equations. To overcome this difficulty, we start by simulating a large database of values for the treatment $T$ and for all the counterfactual mediators $M^k(t)$, and outcomes $Y(t,M^1(t_1),\dots,M^K(t_K))$, see Table \ref{sim_table} for an example. Then we simply calculate the indirect effects $\delta^k(t)$ and $\delta^Z(t)$ and the direct effect $\zeta(t)$ as means, according to the definitions given in the section 3.\ref{definitionmultiple}. The large size of the dataset guarantees that these Monte-Carlo estimates can be taken as the true values. In this study we generate a dataset of $10^6$ observations, so that the estimate error is as small as the $0.2\%$ of the standard deviation of the effect of interest.

\medskip

 In order to obtain a subset of observations on which to test estimation methods, we sample $N$ individuals (i.e. rows) $i=1,\ldots,N$ and for each of them we select only the values $Y(T_i,Z_i(T_i))$ and $Z_i(T_i)$ corresponding to the specific value of $T_i$. More precisely: 
\begin{itemize}
\item if $T_i = 0$ we take $Z_i=(M^1_i,\ldots,M^k_i)=(M^1_i(0),\ldots,M^k_i(0))=Z_i(0)$  and $Y_i=Y_i(0,Z_i(0))$,
\item if $T_i = 1$ we take $Z_i=(M^1_i,\ldots,M^k_i)=(M^1_i(1),\ldots,M^k_i(1))=Z_i(1)$  and $Y_i=Y_i(1,Z_i(1))$.
\end{itemize}
Tables \ref{sim_table} and \ref{2Mta} illustrate the simulation procedure.

\medskip


\medskip

 We consider two causal simulation models, described in appendix \ref{model}, accounting for two types of outcome (continuous and logit binary), and two settings with two continuous causally unrelated mediators. Uncausally correlated mediators, Figure \ref{fig1c}, are simulated from a bivariate normal distribution with fixed covariance matrix.

\medskip 

 For each simulation model, we estimate the different effects of interest by means of the general algorithm for multiple mediators described above in section 3.\ref{ssec:algo}. We compare our estimates with both the true values and the estimates of two simple analyses (one for each mediator) obtained with the \texttt{mediation} package. Because in general $\delta^k(1)\neq\delta^k(0)$ and $\zeta(1)\neq\zeta(0)$, for the sake of simplicity  we focus on average effects such as $\delta=(\delta(0)+\delta(1))/2$ and $\zeta=(\zeta(1)+\zeta(0))/2$. Note that for continuous outcome and in absence of interaction between treatment and mediators, Corollaries \ref{cor1} and \ref{cor1J} imply that $\delta^k(1)=\delta^k(0)$ and $\zeta(1)=\zeta(0)$. For each mediator, we also show the proportion mediated $PM^k=\delta^k/\tau$.

\begin{table}[htp]
\scriptsize
\centering

\begin{tabular}{cccccccccc}
\toprule
\(T\) & \(M(0)\) & \(M(1)\) & \(W(0)\) & \(W(1)\) & \(Y(1,M(1),W(1))\) &
\(Y(1,M(1),W(0))\) & \(Y(1,M(0),W(1))\)    \tabularnewline
\midrule

\textbf{0} & \textbf{0.28} & 1.08 & \textbf{0.53} & 1.43 & 2.42 & 1.79 & 1.94  \tabularnewline
\textbf{0} & \textbf{0.42} & 1.22 & \textbf{-1.80} & -0.90 & 1.41 & 0.78 & 0.93 \tabularnewline
\textbf{1} & 0.63 & \textbf{1.43} & 0.03 & \textbf{0.93} & \textbf{1.87} & 1.24 & 1.39  \tabularnewline
\textbf{1} & 0.75 & \textbf{1.55} & 2.24 & \textbf{3.14} & \textbf{2.95} & 2.32 & 2.47  \tabularnewline
\bottomrule
\end{tabular}

\begin{tabular}{cccccccc}
\toprule
 \(Y(0,M(1),W(1))\) &  \(Y(1,M(0),W(0))\) & \(Y(0,M(1),W(0))\) &
\(Y(0,M(0),W(1))\) & \(Y(0,M(0),W(0))\)\tabularnewline
\midrule

2.02 & 1.31 & 1.39 & 1.54 & \textbf{0.91}\tabularnewline
1.01	& 0.30 &  0.38 & 0.53 & \textbf{-0.09}\tabularnewline
1.47	& 0.76 &  0.84 & 0.99 & 0.36\tabularnewline
2.55 & 1.84 &  1.92 & 2.07 & 1.44\tabularnewline
\bottomrule
\end{tabular}

\caption{Simulated counterfactuals with two independent mediators.} 
\label{sim_table}
\end{table}
\begin{table}[htp]
\centering
\begin{tabular}{cccc}
\toprule
\(T\) & \(M\) & \(W\) & \(Y\)\tabularnewline
\midrule
0 & 0.28 & 0.53 & 0.91\tabularnewline
0 & 0.42 & -1.80 & -0.09\tabularnewline
1 & 1.43 & 0.93 & 1.97\tabularnewline
1 & 1.55 & 3.14 & 2.95\tabularnewline
\bottomrule
\end{tabular}
\caption{Simulated observed data with two independent mediators. Observations were extracted from Table \ref{sim_table}.}
\label{2Mta}
\end{table}

\subsection{Limitations of repeated simple analyses when the common cause of mediators is not measured}\label{imailimit}

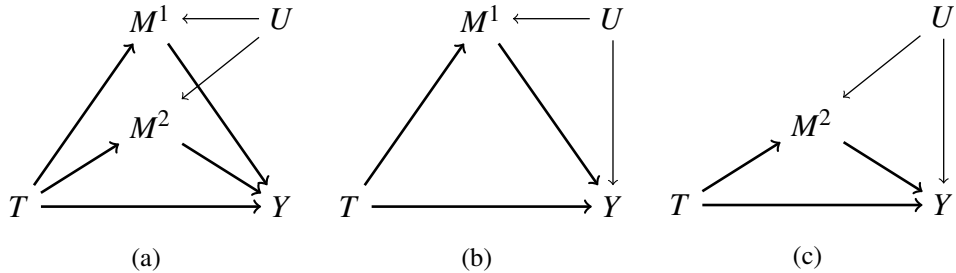
\begin{figure}[htp]
\begin{center}
\minipage{0.3\textwidth}
\subfigure[]{\label{fig5a}

\begin{tikzpicture}
\node (t) at (0,0) {$T$};
\node (y) at (3.5,0) {$Y$};
\node (m) at (1.75,2.5) {$M^1$};
\node (w) at (1.75,1.1) {$M^2$};
\node (u) at (3.5,2.5) {$U$};

\draw[->,line width= 0.5] (u) --  (m);
\draw[->,line width= 0.5] (u) --  (w);
\draw[->,line width= 1] (t) --  (y);
\draw [->,line width= 1] (m) -- (y);
\draw [->,line width= 1] (t) -- (m);
\draw [->, line width= 1] (w) -- (y);
\draw [->,line width= 1] (t) -- (w);
\end{tikzpicture}
}
\endminipage \hspace{0.1cm}
\minipage{0.3\textwidth}
\subfigure[]{\label{fig5b}
\begin{tikzpicture}
\node (t3) at (0,0) {$T$};
\node (y3) at (3.5,0) {$Y$};
\node (m3) at (1.75,2.5) {$M^1$};
\node (u) at (3.5,2.5) {$U$};



\draw[->,line width= 1] (t3) --  (y3);
\draw [->,line width= 1] (m3) -- (y3);
\draw [->,line width= 1] (t3) -- (m3);
\draw [->, line width= 0.5] (u) -- (m3);
\draw [->, line width= 0.5] (u) -- (y3);

\end{tikzpicture}
}
\endminipage\hspace{0.1cm}
\minipage{0.3\textwidth}
\subfigure[]{\label{fig5c}
\begin{tikzpicture}
\node (t3) at (0,0) {$T$};
\node (y3) at (3.5,0) {$Y$};
\node (w3) at (1.75,1.1) {$M^2$};
\node (u) at (3.5,2.5) {$U$};



\draw[->,line width= 1] (t3) --  (y3);
\draw [->,line width= 0.5] (u) -- (w3);
\draw [->, line width= 0.5] (u) -- (y3);
\draw [->, line width= 1] (w3) -- (y3);
\draw [->,line width= 1] (t3) -- (w3);

\end{tikzpicture}
}
\endminipage
\end{center}
\caption{Multiple and simple mediation analyses, $U$ observed. Data are simulated according to the model in (a).}\label{underU}
\end{figure}

Data are generated under the model described in Figure \ref{fig5a}, where the dependence between the two mediators is induced by the pre-treatment variable $U$. More specifically, variables are simulated according to the following distributions:

\begin{itemize}

\item $T$ follows a Bernoulli distribution $\mathcal{B}(0.3)$

\item $U$ follows a normal distribution $\mathcal{N}(0,1)$
\item The conditional distribution of the two counterfactual mediators given $T$ and $U$ are

\begin{eqnarray*}
M^1(t)|T=t,U=u & \sim & \mathcal{N}\left(1+ 4t + 2  u,1\right) \\
M^2(t)|T=t,U=u  & \sim & \mathcal{N}\left(2 +6  t+ 3  u,1\right)
\end{eqnarray*}
\item The counterfactual outcome follows the normal distribution
\begin{eqnarray*}
Y\left(t,M^1(t'),M^2(t'')\right)  & \sim &  \mathcal{N}(1+10  t + 5  M^1(t') + 4  M^2(t'') ,1).
\end{eqnarray*}
\end{itemize}

 Note that the correlation between the two mediators conditionally on the treatment (and not on $U$, Figure \ref{fig1c}), is equal to 0.7.

 When we have two causally independent mediators and $U$ is observed, the approach by \cite{imai_identification_2013} is to perform two simple analyses as in Figure \ref{fig5b} and \ref{fig5c}. 

%
%
%
%
%
%
%
%
%

\medskip

 However, when $U$ is unobserved, the situation is like in Figure \ref{fig1c} with mediators showing residual correlation. In this case, conducting separate simple analyses is not appropriate because Sequential Ignorability assumptions (\ref{ISB2}) and (\ref{ISB3}) are violated (\cite{imai_identification_2013}).

\medskip 

 Here we illustrate this problem through simulations. For comparison purposes we also add results obtained with our method for multiple analysis.

\medskip

\begin{table}[htp]
\centering
\begin{tabular}{|c|c|c|c|c|}
\hline
Effects 			 &Value& Simple Analysis $M^1$ & Simple Analysis $M^2$	& Multiple Analysis 					\\
\hline
$\delta^Z$ & 44 &  &  & 44.44 [43.32 ; 45.52] \\ 
$PM^Z$ & 0.81 &  &  & 0.81  [0.81 ; 0.82] \\ 
$\delta^1$ & 20 & 19.38 [17.72 ; 21.06] &  & 20.55 [19.60 ; 21.45] \\ 
$PM^1$ & 0.37 & 0.36  [0.33 ; 0.39] &  & 0.38 [0.36 ; 0.40] \\ 
$\delta^2$ & 24 &  & 21.63 [18.79 ; 24.60] & 23.89 [23.10 ; 24.66] \\ 
$PM^2$ & 0.44 &  & 0.40  [0.34 ; 0.45] & 0.44  [0.42 ; 0.45] \\ 
$\zeta$ & 10 & 35.00 [33.47 ; 36.66] & 32.80 [29.72 ; 35.65] & 9.99 [9.76 ; 10.22] \\ 
$\tau$ & 54 & 54.37 [53.39 ; 55.46] & 54.42 [53.28 ; 55.50] & 54.43 [53.25 ; 55.52]\\ 
   \hline
\end{tabular}
\caption{Adjusting for $U$ when all variables in Figure \ref{fig5a} are observed.}\label{tab:underU}
\end{table}

\begin{table}[htp]
\centering
\begin{tabular}{|c|c|c|c|c|}
\hline
Effects 			 &Value& Simple Analysis $M^1$ & Simple Analysis $M^2$ & Multiple Analysis 					\\
\hline
$\delta^Z$ & 44 &  &  & 43.25 [40.2447 ; 46.25] \\ 
$PM^Z$ & 0.81 &  &  & 0.81 [0.80 ; 0.82] \\ 
$\delta^1$ & 20 & 38.45 [34.53 ; 42.34] &  & 20.00 [18.04 ; 22.19]\\ 
$PM^1$ & 0.37 & 0.72  [0.68 ; 0.75] &  &  0.37 [0.33 ; 0.42] \\ 
$\delta^2$ & 24 &  & 40.86 [37.03 ; 44.98] & 23.24 [20.92 ; 25.51] \\ 
$PM^2$ & 0.44 &  & 0.76 [0.73 ; 0.79] & 0.43 [0.38 ; 0.48] \\ 
$\zeta$ & 10 & 14.75 [13.19 ; 16.42] & 12.46 [10.89 ; 13.89] & 9.98 [9.75 ; 10.22] \\ 
$\tau$ & 54 & 53.21 [49.20 ; 57.06] & 53.32 [49.51 ; 57.21] & 53.23 [50.20 ; 56.25] \\ 
   \hline
\end{tabular}
\caption{Not adjusting for $U$: data are generated  as in Figure \ref{fig5a} but analyzed as if $U$ was unobserved.}
\label{tab:noU}
\end{table}

\medskip

 As expected, Tables \ref{tab:underU} and \ref{tab:noU} show that simple analyses adjusted for $U$ give precise and accurate estimates of indirect effects (but obviously not of the direct effect), while they give biased estimates when U is not taken into account. On the contrary, our method gives precise and accurate estimates of all effects with or without taking into account $U$, showing that it is still possible to conduct a mediation analysis to estimate all effects even when $U$ is unobserved. 
 
\medskip

In the following subsection we suppose that $U$ is unobserved, as it is often the case in practical situations.

\subsection{Empirical study of the properties of the proposed estimators}
\label{ss:empirical_study}

 The previous section illustrated our method on a single simulation run. In this section, we perform a simulation-based study to assess the properties of the of proposed estimators. More specifically, we calculate bias, confidence interval coverage probability, mean square error (MSE) and variance of our estimators as means over $200$ simulation runs for each considered parameter setting. We compare the estimates of several simple analyses, one for each mediator, to the estimates obtained with our multiple mediation analysis for different correlation levels. This comparison is done on the two models described in appendix \ref{model}. Both have continuous mediators but the outcome is continuous for model 1 and binary logistic for model 2.

\medskip

\begin{landscape}
\begin{figure}
\vspace{-3cm}
\centering
\includegraphics[height=0.3\textwidth]{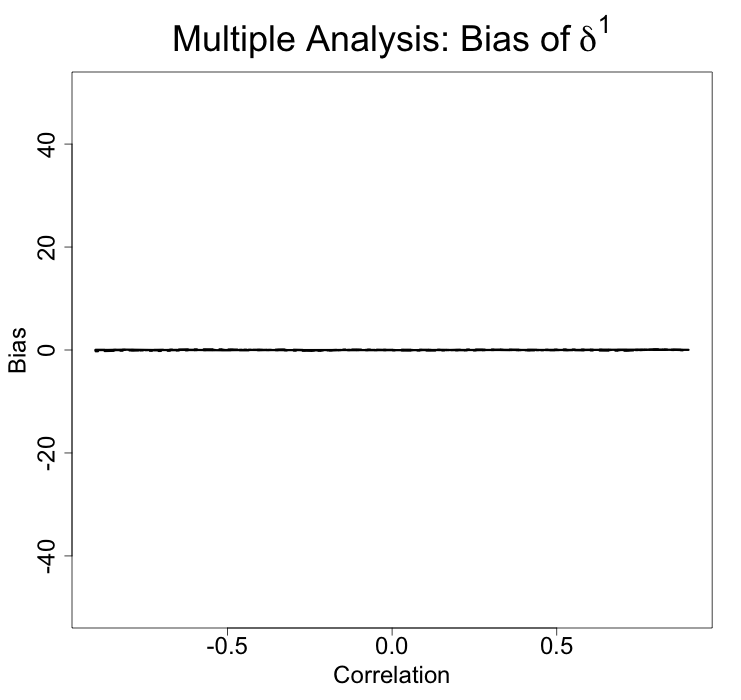}
\includegraphics[height=0.3\textwidth]{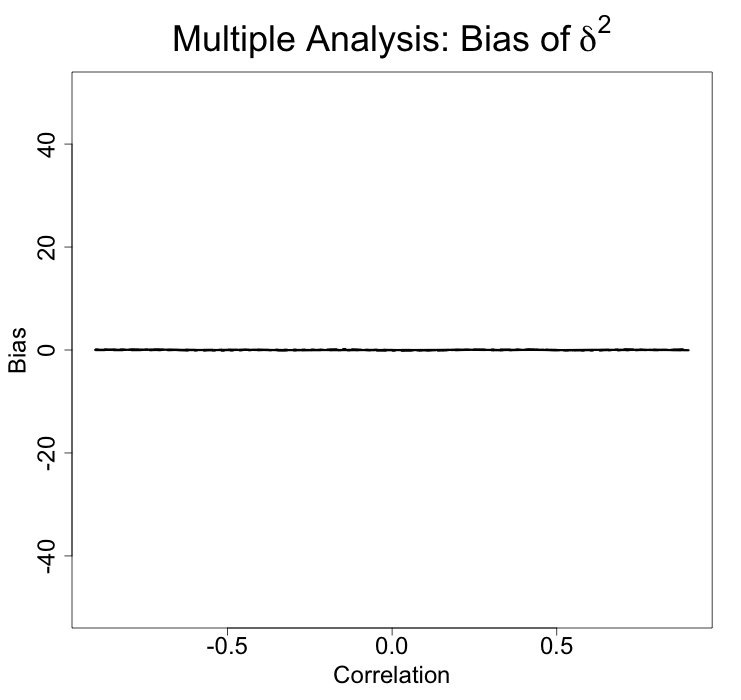}
\includegraphics[height=0.3\textwidth]{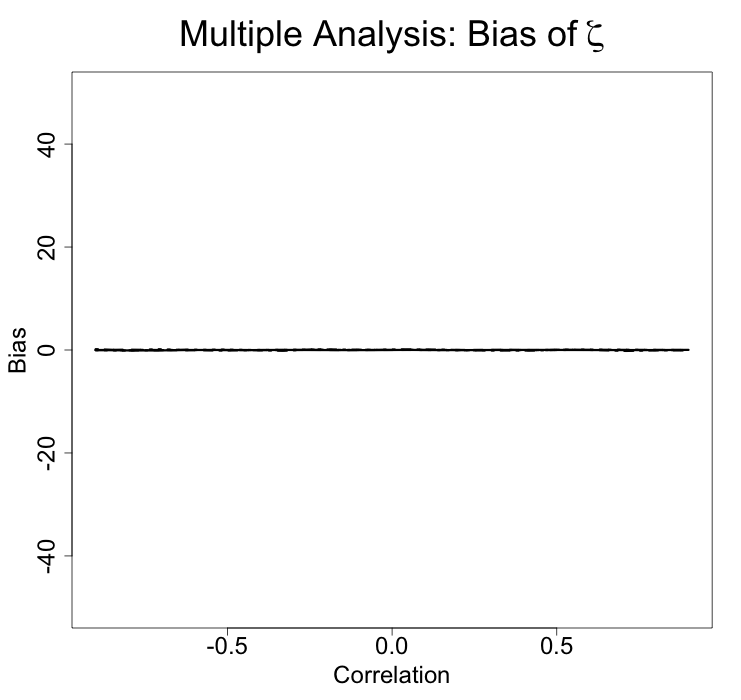}
\includegraphics[height=0.3\textwidth]{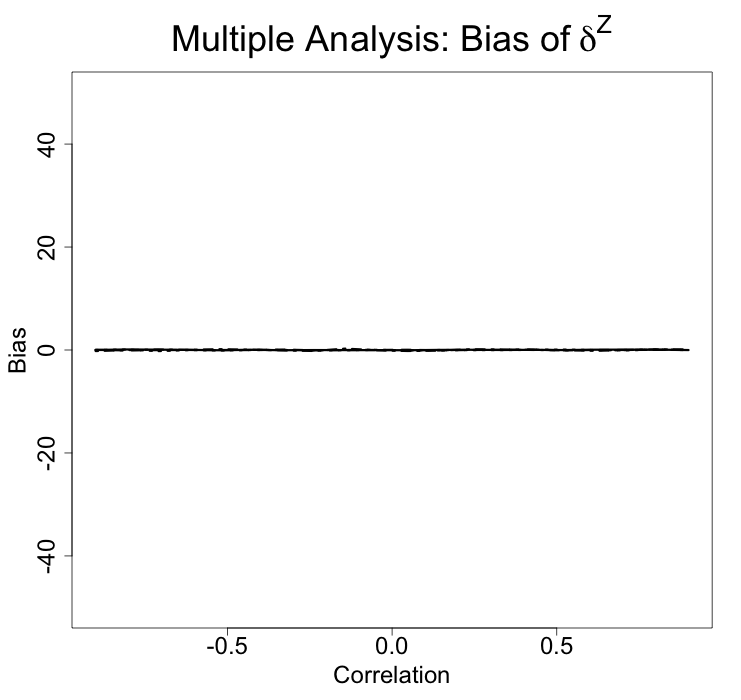}
\includegraphics[height=0.3\textwidth]{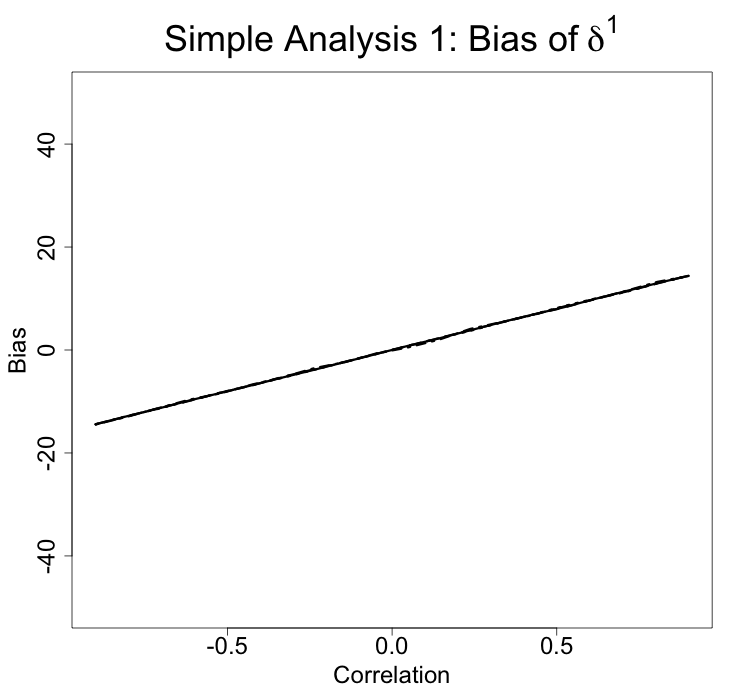}
\includegraphics[height=0.3\textwidth]{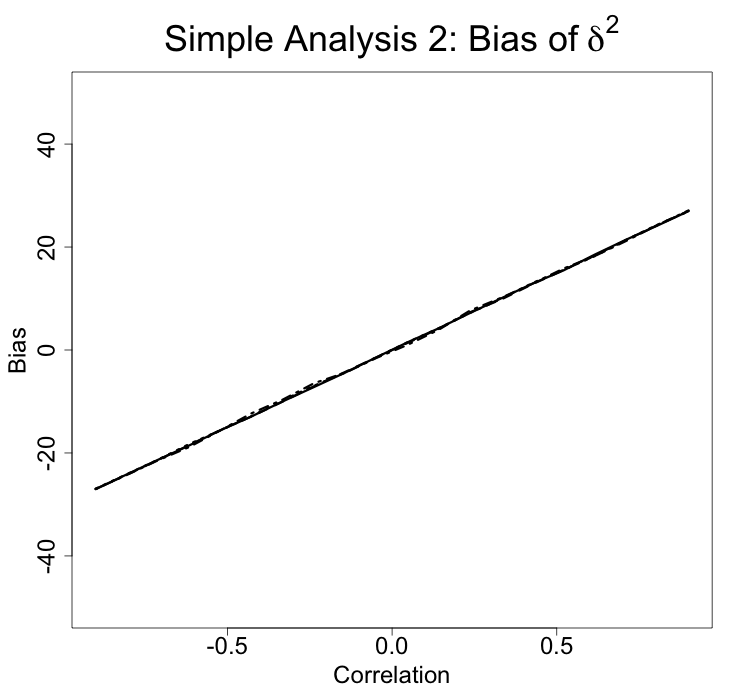}
\includegraphics[height=0.3\textwidth]{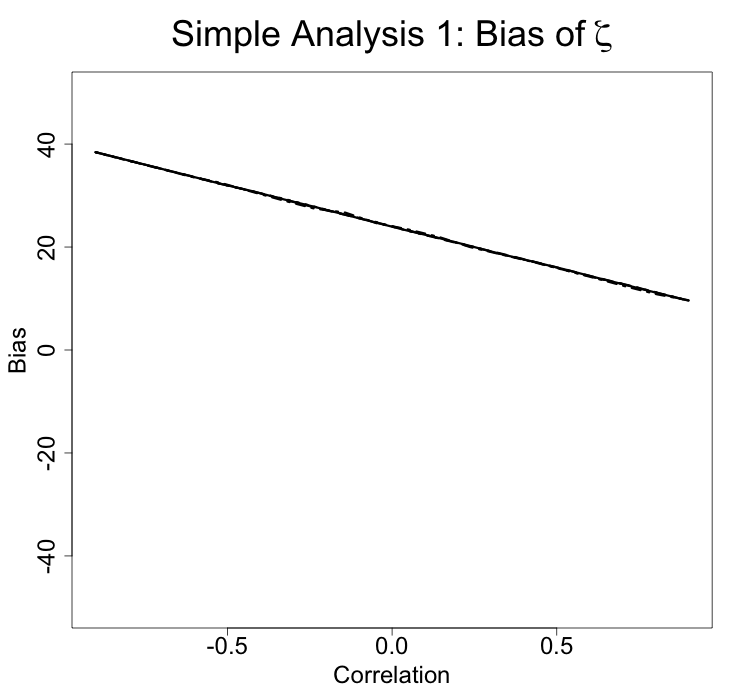}
\includegraphics[height=0.3\textwidth]{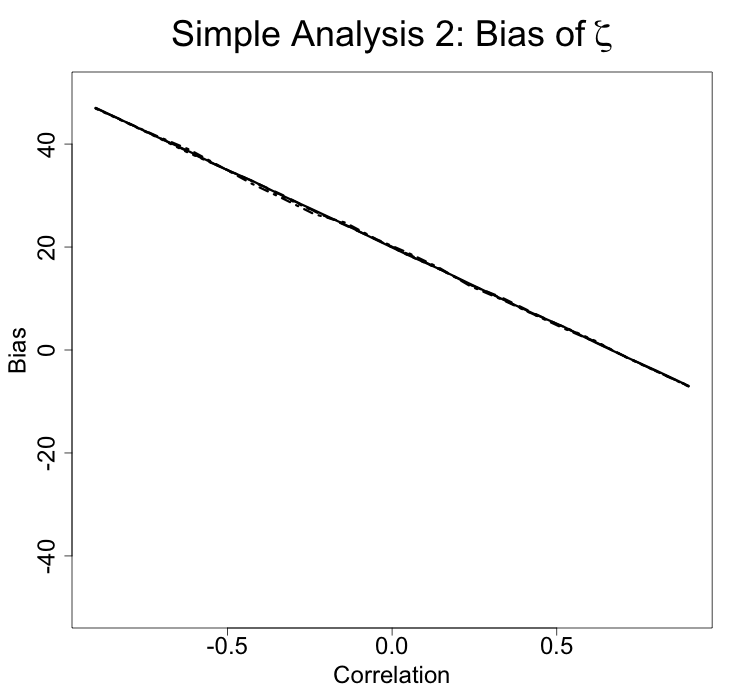}
\includegraphics[height=0.3\textwidth]{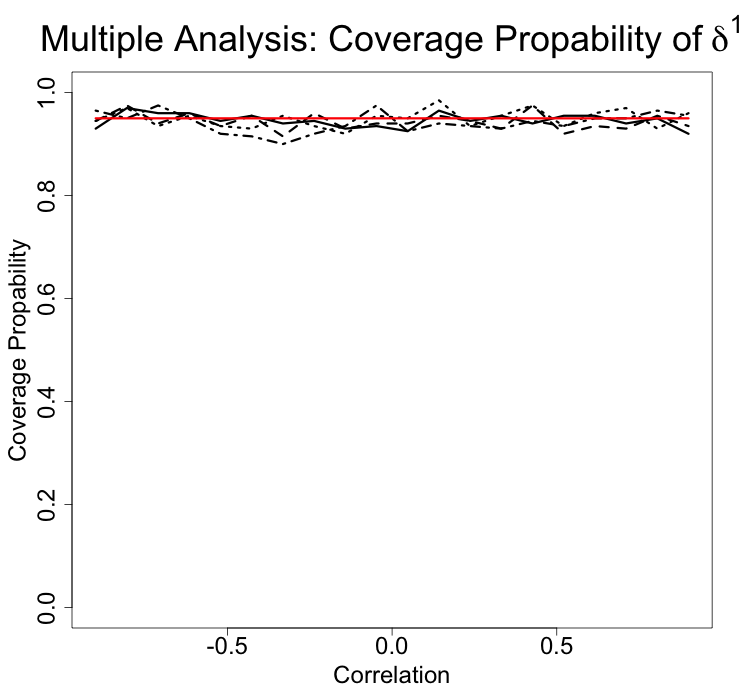}
\includegraphics[height=0.3\textwidth]{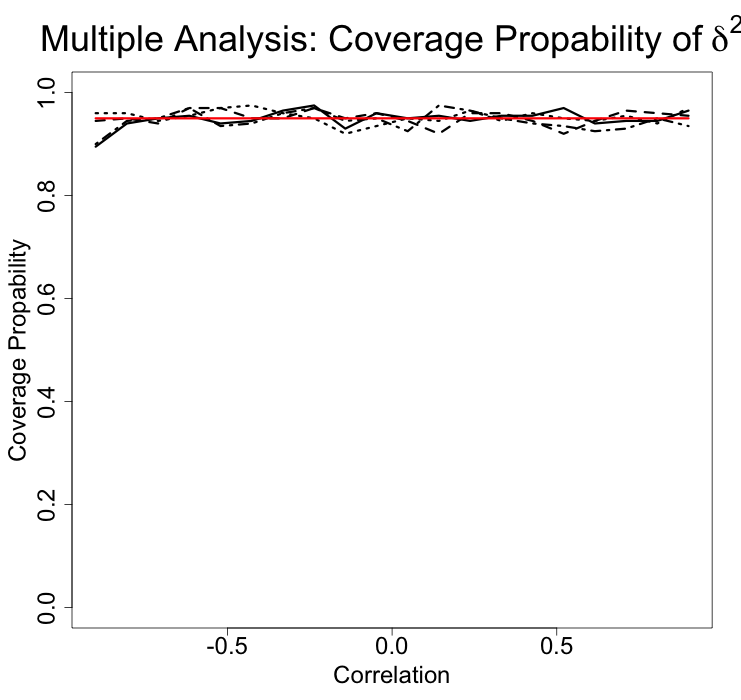}
\includegraphics[height=0.3\textwidth]{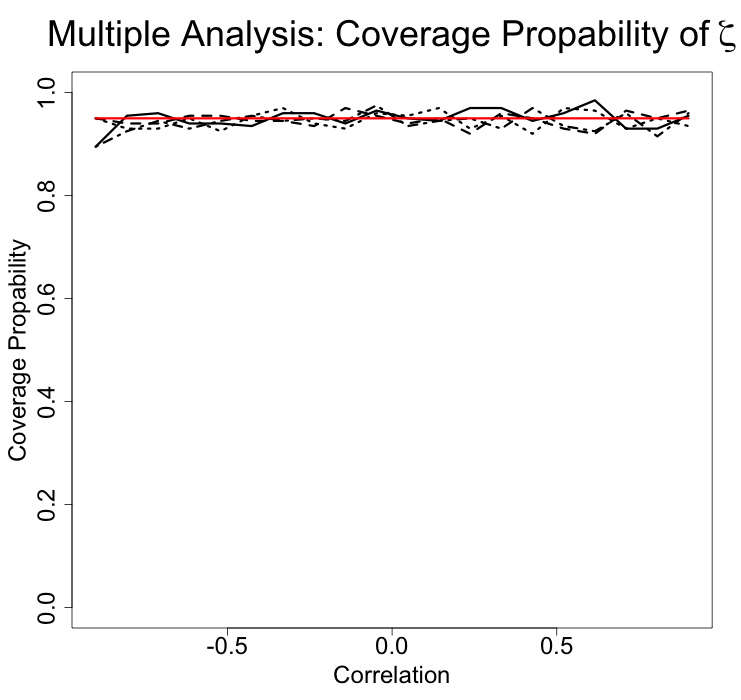}
\includegraphics[height=0.3\textwidth]{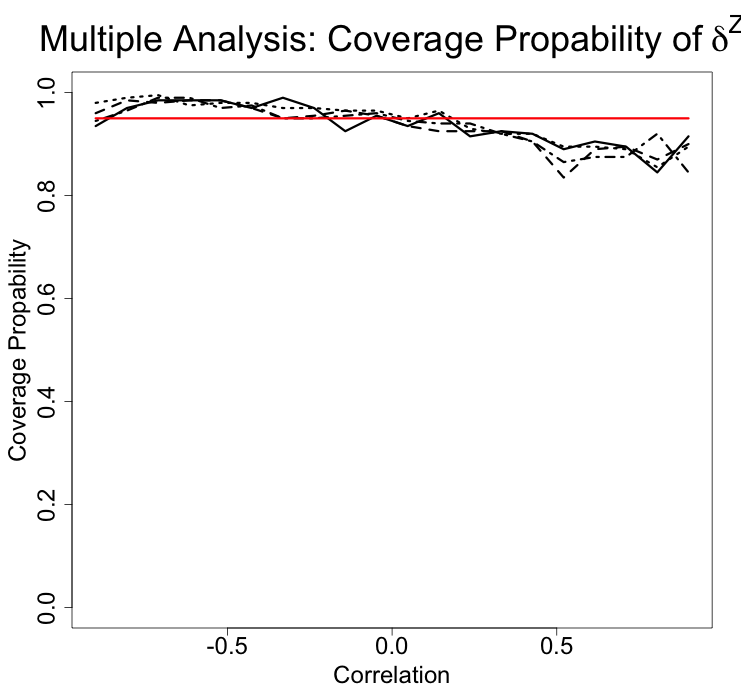}
\includegraphics[height=0.3\textwidth]{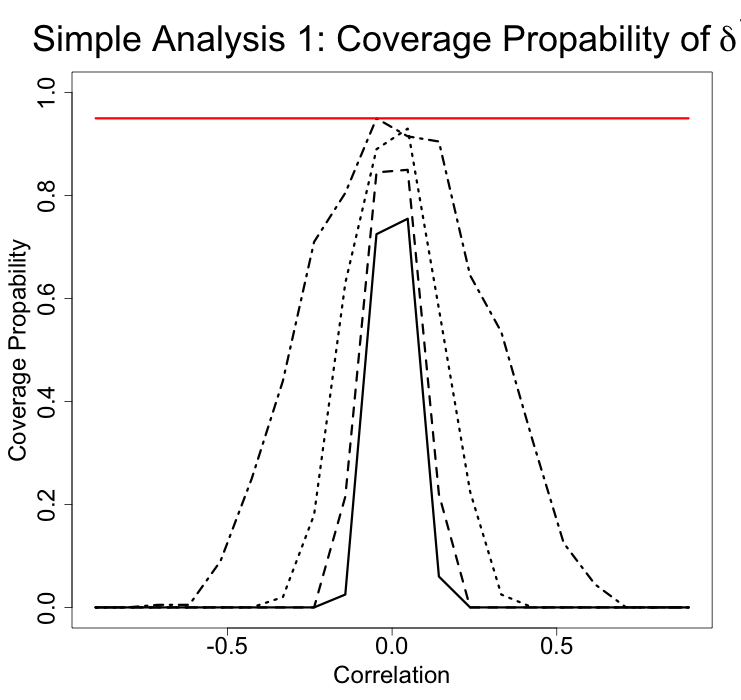}
\includegraphics[height=0.3\textwidth]{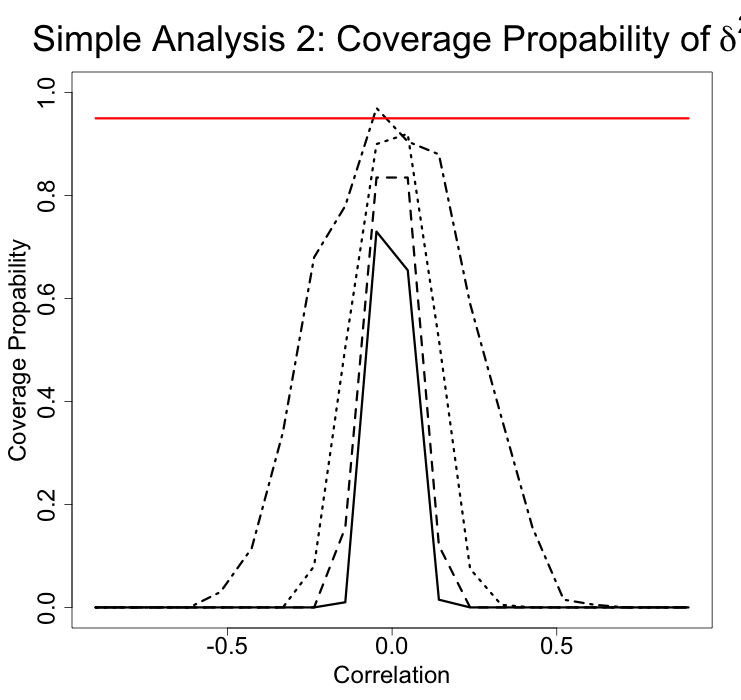}
\includegraphics[height=0.3\textwidth]{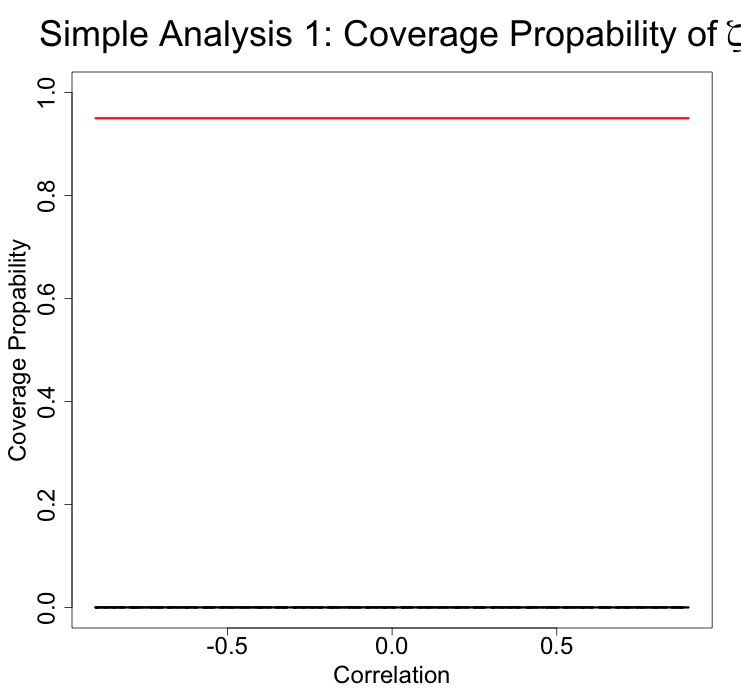}
\includegraphics[height=0.3\textwidth]{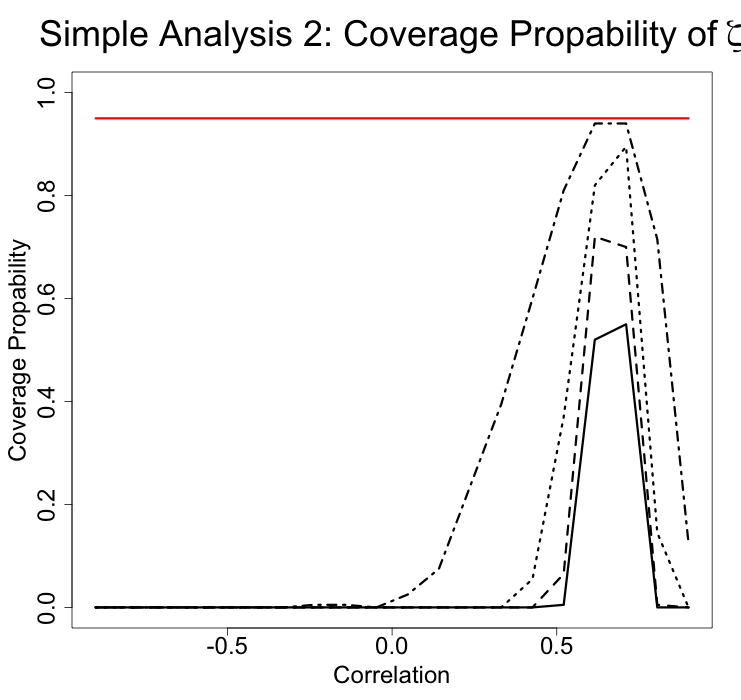}
\includegraphics[height=0.05\textwidth]{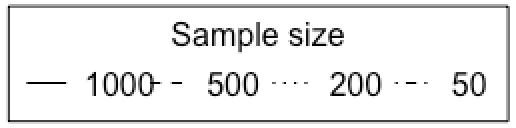}
\caption{Bias and coverage probability of the estimators of indirect and direct effects when the correlation between mediators varies  in model 1. The bias formula used here is $Bias = \Theta-\mathbb{E}\left[\widehat{\Theta}\right]$. The first two rows show the bias (multiple analysis in row 1 and simples analyses in row 2) and the last two show the coverage probability (multiple analysis in row 3 and simples analyses in row 4).}\label{resultsim}
\end{figure}
\end{landscape}

  Simulations are run for model 1 for different values of correlation between the mediators and increasing observed data sample size (50, 200, 500 and 1000).  Results  for bias and coverage probability can be seen in Figure \ref{resultsim}. This figure clearly shows that our approach allows an unbiased estimation, contrary to the simple analyses, for both direct and indirect effects. Interestingly the sum of the bias of the direct effect and of the indirect effect of a given mediator estimated by a simple analysis is constant and corresponds to the opposite of the indirect effect of the mediator not taken into account, as shown by Figure \ref{sumbias} in the Appendix.

The empirical 95\% confidence interval given by our method contains the real value in approximatively 95\% of the runs, for both indirect and direct effects and whatever the correlation between the mediators. On the contrary, simple analyses obtains fair coverages only when the correlation is almost null.
Figure \ref{resultsimother} in the Appendix shows that our estimators have low variance and low MSE for sample sizes larger than $200$.
\medskip

  Simulations were also run for model 2 for different values of correlation between the mediators 1000 observational data. As illustrated by Figure \ref{resultsimeff} in the Appendix, the results for bias, coverage probability, variance and MSE confirms that our estimators are unbiased and have low variance and the expected coverage probability, thus outperforming simple analyses.

\medskip

\medskip

\medskip

\medskip


\section{Application} \label{application}

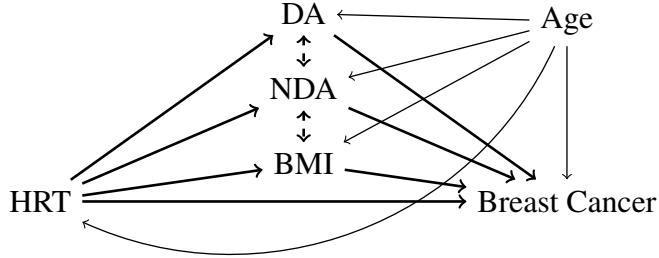
\begin{figure}[htp]
\begin{center}
\begin{tikzpicture}
\node (t) at (0,0) {HRT};
\node (y) at (7,0) {Breast Cancer};
\node (m1) at (3.5,2.5) {DA};
\node (m2) at (3.5,1.5) {NDA};
\node (m3) at (3.5,0.5) {BMI};
\node (x) at (7,2.4) {Age};
 
\draw[->, line width= 1] (t) --  (y);
\draw [->, line width= 1] (m1) -- (y);
\draw [->, line width= 1] (t) -- (m1);
\draw [->, line width= 1] (m2) -- (y);
\draw [->, line width= 1] (t) -- (m2);
\draw [->, line width= 1] (m3) -- (y);
\draw [->, line width= 1] (t) -- (m3);
\draw [<->, dashed, line width =1] (m1) -- (m2);
\draw [<->, dashed, line width =1] (m2) -- (m3);
\draw [->, line width= 0.5] (x) to [bend left=45] (t);
\draw [->, line width= 0.5] (x) -- (y);
\draw [->, line width= 0.5] (x) -- (m1);
\draw [->, line width= 0.5] (x) -- (m2);
\draw [->, line width= 0.5] (x) -- (m3);
\end{tikzpicture}
\end{center}
\caption{Causal diagram for the application.}\label{cormed}
\end{figure}

We applied our method to a real data set to estimate the amount of causal effect of hormone replacement therapy (HRT) on breast cancer (BC) risk that is mediated by mammographic density (MD) - specifically dense area (DA) and non-dense area (NDA) -  and body mass index (BMI)  in postmenopausal women. The data come from the E3N French cohort study (\cite{clavel-chapelon_cohort_2015}). HRT, prescribed to relief menopausal symptoms, consists in providing women with hormones whose production naturally decreases with menopause (\cite{miller_update_2017}). One of the consequence of taking HRT is that women do not experience the decrease of DA, the increase of NDA and the increase of BMI typically occurring at menopause
(\cite{mctiernan_estrogen-plus-progestin_2005}). HRT use has been since long recognised to be a risk factor for BC (\cite{kim_menopausal_2018}). Independent BC risk factors are also high postmenopausal BMI and high per age and per BMI MD (\cite{baglietto_associations_2014, maskarinec_tumor_2017}). In order to better understand the mutual relationship between HRT, MD and BMI in BC carcinogenesis, it is important to determine whether and eventually to which extent the effect of HRT on BC risk is due to its action on MD and BMI (mediated effect) and to which extent it is independent of MD and BMI (direct effect).

The continuous variables were normalised using the Box-Cox likelihood-like approach (\cite{boxcox_1964}),
$t(M)  = \dfrac{M^\lambda-1}{\lambda}$, with $\lambda$ equal to 0.38, 0.34 and -1.19 for DA, NDA and BMI respectively. HRT was treated as a dichotomous variable whose levels were never versus ever users (past or current).

\medskip

\subsection{Regression models}

\begin{table}[htp]
\scriptsize
\centering
\begin{tabular}{l|l|ccccc}
\hline
Model && HRT & AGE & DA & NDA & BMI \\
\hline
1 &DA &10.68 &-0.36&-&-&- \\
2 & NDA & -5.72&0.60&-&-&- \\
3 & BMI & -0.73& 0.04&-&-&- \\
4.a & BC & 0.49&0.003&-&-&-  \\
4.b & BC & 0.39&0.01&0.01& -0.01&0.10 \\
\hline
\end{tabular}
\caption{Estimation of the regression coefficients. The second column contains the variables explained by the explanatory variables in the first line. For example $DA \sim 10.68 HRT - 0.36 AGE$. Note that we have a logistic regression for BC.}\label{regreappli}
\end{table}

In preparation to our mediation analysis, we regressed each mediator on HRT and AGE (Table \ref{regreappli}, models 1, 2 and 3 respectively) and BC on HRT and AGE with or without conditioning on the three mediators (respectively models 4a, 4b). As expected, HRT ever users had significantly higher values of DA and significantly lower of NDA and BMI (Table \ref{regreappli}); DA and BMI were positively associated with BC risk, whereas NDA was negatively associated with risk (Table \ref{regreappli}). HRT was positively associated with BC risk and the association decreased of the 20\% in the log-OR scale when accounting for DA, NDA and BMI into the model (Table \ref{regreappli} models 4a and 4b). Note that after adjusting for HRT and Age the residuals correlation between DA and BMI, NDA and BMI and DA and NDA are -0.04, -0.22 and 0.60 respectively.

\medskip

\subsection{Multiple mediation analysis}

We applied our method with models 1 2, 3 and 4.b from Table \ref{regreappli} to estimate the causal mediated effect due to all mediators and the causal mediated effect due to each of them when accounting for their mutual correlation. As shown in Table \ref{appli2} the causal mediated effects due to DA and NDA were positive, whereas the causal mediated effect due to BMI was negative; this resulted in a proportion of the total mediated effect of 22\% (95\% CI: 1\% to 63\%).
Our finding that the effect of HRT is partially mediated by MD is consistent with previous reports in the literature (\cite{rice_does_2018, azam_hormone_2018}).

\begin{table}[htp]
\scriptsize
\centering
\begin{tabular}{|llc|}
\hline
& Estimate & $95\%$IC  \\
			 \hline
$\hat{\delta}^{DA}$ & 0.0251 &[0.0121 ; 0.0414] \\ 
$\hat{\delta}^{NDA}$ & 0.0122 & [0.0019 ; 0.0255] \\ 
$\hat{\delta}^{BMI}$ & -0.0149 &[-0.0305 ; -0.0038] \\ 
$\hat{\delta^{Z}}$ & 0.0224 & [0.0014 ; 0.0439]\\ 
$\hat{PM^{Z}}$ & 0.2154 & [0.0119 ; 0.6302]\\
$\hat{\zeta}$ & 0.0800 & [0.0160 ; 0.1471] \\
$\hat{\tau}$ & 0.1024 & [0.0358 ; 0.1660]\\
\hline
\end{tabular}
\caption{Multiple mediation analysis for $T\in\{0,1\}$ (i.e. never vs  former/current HRT users).}\label{appli2}
\end{table}

\section{Discussion}

This article adresses the problem of estimating direct and indirect effects, including indirect effects through individual mediators, in the framework of multiple mediation with uncausally related mediators. Theoretical work of Shpitser and coauthors proved that in presence of latent variables not all mediation quantities are identified (\cite{shpitser_counterfactual_2013, shpitser_indentification_2018}). In particular, in presence of a latent common cause between the mediators, indirect effects trough individual mediators cannot be expressed as functions of the observable data only. On the other hand, a common practice in multiple mediation is to perform several simple mediation analyses, one for each mediator, despite the introduction of a bias.\\
 We define a set of hypotheses, called SIMMA, under which we express the direct and the joint indirect effect as functions of observed variables and the indirect effect through individual mediators in terms of both observed and counterfactual variables. Coupled to a choice of model and the quasi-Bayesian algorithm developed by \cite{imai_general_2010}, the latter formula gives an estimation method for the individual indirect effects. Note that we restricted ourselves to models with the additional hypothesis that the correlation between counterfactual mediators is the same whatever the treatment governing them. The development of sensitivity analysis methods to test the robustness of our results to the violation of this hypothesis would require the parametrization of our formulas in terms of the correlation between potential mediators under specific parametric models. We leave this important perspective, together with the developpment of a sensitivity analysis for assessing the robustness to SIMMA violations, for a future work.\\
 
 The method is implemented in R. Currently our program makes it possible to work with parametric models with continuous mediators and continuous or binary outcomes. A package will soon be published, possibly extending the current framework to other kind of models (e.g. for categorical mediators) and including methods for sensitivity analysis.

%

We applied our R program to validate the proposed method empirically. This simulation study shows that our method provides an unbiased estimate of the direct effect, while, as expected, estimates obtained by running simple mediation analyses one mediator at the time are biased, even in the case of independent mediators. Moreover, when mediators share an unobserved common cause, we show that our multiple analysis provide estimates of the direct effects through individual mediators that are less biased than the ones obtained from simple analyses one mediator at the time. The reason behind this improvement, is that our method, by considering the joint law of the mediators conditionally on the treatment and the law of the outcome conditionally on all the mediators, automatically takes into account the influence that the unobserved common cause $U$ has on the mediators and the outcome. On the contrary, doing a simple analysis one mediator at the time is not appropriate in this setting because $U$ confounds the relationship between each mediator and the outcome. Moreover, we show empirically that, contrary to repeated simple analyses, the proposed quasi-bayesian algorithm provides confidence intervals with the expected coverage property.

Repeated individual mediator analyses are still a popular approach despite a growing literature warning about its limitations. Indeed, the presence of an unobserved common cause for the mediators is not the only situation in which such an approach is problematic. \cite{vanderweele_mediation_2014} observed that, even when mediators are uncausally related, it is not possible to decompose the joint indirect effect in the sum of individual indirect effects when their effect on the outcome is characterized by an interaction in the additive scale, a situation we excluded in our theoretical results. In this situation, \cite{taguri_causal_2015} provided a three way decomposition of the joint indirect effect into individual natural indirect effects and an interactive effect. Interestingly, the assumptions required to show the identifiability of all the terms in this decomposition are the same as ours, with the only important difference that potential mediators are assumed to be conditionally independent given all observed covariates. More recently, \cite{bellavia2017decomposition} provided a decomposition of the total effect in the more general situation with both mediator-mediator and mediators-outcome interactions.

Another important setting where repeating simple analyses is the wrong approach to multiple mediation is when mediators are causally ordered as in Fig. \ref{fig1b}. In this situation, considering the vector of intermediate variables as one mediator and conducting a simple analysis will correctly estimate the joint indirect effect and the direct effect. However the former joint indirect effect is not equal to the sum of the individual indirect effect, each estimated with a simple analysis, because some paths are counted twice and the effect mediated by $W$ is biased by $M$ which acts as a posttreatment confounder of the $W-Y$ relationship. More generally, unless strong conditions hold it is not possible to identify all specific paths (\cite{Avin2005,daniel_causal_2015}). \cite{vanderweele_mediation_2014} introduced a sequential approach to identify the joint indirect effect, the direct effect, the effect mediated by $M$ and the effect mediated by $W$ but not $M$. The different steps in this strategy can be implemented using \texttt{medflex}, a recently introduced R package based on the \textit{natural effect model} and imputation or weighting methods (\cite{steen2017medflex}). An alternative approach based on linear structural equations with varying coefficients was discussed by \cite{imai_identification_2013} and implemented in the \texttt{mediation} package. \cite{nguyen2015practical} presented a method based on the Inverse Odds Ratio Weighting (IOWR) approach introduced by \cite{tchetgen2013inverse}. This method is very flexible as it accommodates generalized linear models, quantile regression and survival models for the outcome and multiple continuous or categorical mediators, however it does not allow to estimate the indirect effect through individual mediators, but only the joint indirect effect.

We conclude this brief overview of the literature around multiple mediation by underlining that our framework deals with \textit{natural} direct and indirect effects. \cite{vansteelandt2017interventional} recently introduced so-called \textit{interventional} direct and path specific indirect effects that do add up to the total effect and are identifiable even when the mediators share unmeasured common causes or the causal dependence between mediators is unknown.

As an illustration of our method, we conducted a multiple mediation analysis on a real dataset from a large cohort to assess the effect of hormone replacement treatment on breast cancer risk through three non-sequential mediators, namely dense mammographic area, nondense area and body mass index. The causal effects that we have estimated and reported can be interpreted as risk differences, that is differences in percentage points. For a binary outcome, it is however often preferred to measure risk changes in terms of odds ratios (OR). In a parallel work in progress aimed at the epidemiological community, we expand on the application of Section \ref{application} and work out a method to compute the causal effects of interest in the OR scale following the definition by \cite{vanderweele_odds_2010}.

\appendix
\section{Link between $\delta^Z$ and $\displaystyle\sum_{k}\delta^{k}$} \label{link}
Even though intuitively it would sound reasonable to think that the indirect effect via the $k$-th mediator $\delta^k$ is the difference between the joint effect $\delta^Z$ and the indirect effect by all other mediators $\eta^k$, we show that this is not true in general.

 We want to express $\delta^Z$ according to $\displaystyle\sum^K_{k=1}\delta^k$. To do so, we start from $\delta^k$:

\begin{eqnarray*}
\delta^k(t) & = & E[Y(t,M^k(1),W^k(t)) - Y(t,M^k(0),W^k(t))]\\
& = & \left\{
  \begin{array}{ll}
E[Y(1,Z(1)) - Y(1,M^k(0),W^k(1))] &  \text{if} \, t=1\\
E[Y(0,M^k(1),W^k(0)) - Y(0,Z(0))] &  \text{if} \, t=0
\end{array}
\right.\\
& = & \left\{
  \begin{array}{ll}
E[\tau + Y(0,Z(0))-Y(1,M^k(0),W^k(1))] &  \text{if} \,t=1\\
E[Y(0,M^k(1),W^k(0)) +\tau - Y(1,Z(1))] &  \text{if} \, t=0
\end{array}
\right. \\
& = & \left\{
  \begin{array}{ll}
E[\tau + Y(1,Z(0))- \zeta(0) -Y(1,M^k(0),W^k(1))] &  \text{if} \,t=1\\
E[Y(0,M^k(1),W^k(0)) + \tau - \zeta(1) - Y(0,Z(1))] &  \text{if} \, t=0
\end{array}
\right. \\
& = & \left\{
  \begin{array}{llr}
E[\delta^Z(1) - Y(1,M^k(0),W^k(1)) + Y(1,Z(0))] &  \text{if} \,t=1 & \footnotemark\\
E[\delta^Z(0) - Y(0,Z(1)) + Y(0,M^k(1),W^k(0))] &  \text{if} \, t=0 &
\end{array}
\right. \\
& = & \delta^Z(t) - \eta^k(t). \\
\end{eqnarray*}
\addtocounter{footnote}{0}
\footnotetext{In fact $\tau=\delta^Z(t)+\zeta(1-t).$}

 $\eta^k$ may be interpreted as the indirect effect by all mediators except the $k$-th, when the treatment is fixed at  $t$ and the $k$-th mediator is set to the value it would have under treatment $1-t$. Summing over the $K$ mediators, we have:

\begin{eqnarray*}
\sum_{k=1}^{K}\delta^k(t) & = & \sum_{k=1}^{K}\left(\delta^Z(t) - \eta^k(t)\right) \\
 & = & K\delta^Z(t) -\sum_{k=1}^{K} \eta^k(t)
\end{eqnarray*}
 Thus joint indirect effect can be rewritten as:
$$\delta^Z(t)  =  \dfrac{\displaystyle\sum_{k=1}^{K}\left(\delta^k(t) +\eta^k(t)\right)}{K}.$$

\section{Assumptions}\label{hypothese}
 According to \cite{imai_identification_2013}, Sequential Ignorability Assumptions in the situation of multiple mediators that are causally unrelated are:

\begin{eqnarray}
\{Y(t,m,w),M(t'),W(t'')\} & \independent & T|X=x,  \label{ISB1}\\
Y(t',m,W(t')) & \independent &M(t)|T,X=x,  \label{ISB2}\\
Y(t',M(t'),w) & \independent & W(t)|T,X=x,\label{ISB3}
\end{eqnarray}

 where $\mathbb{P}(T=t|X=x)>0$ et $\mathbb{P}(M=m,W=w|T=t,X=x)>0$ for all $x, t, t', m, w$.\\

 Furthermore we add these assumptions: 

\begin{eqnarray}
Y(t,m,w) & \independent & \left(M(t'),W(t'')\right)|T,X=x, \,\forall\, t,t',t'',m,w \label{ISB4}\\
 \left(M(t),W(t)\right) & \independent & T|X=x, \,\forall\, t \label{ISB5}
\end{eqnarray}

 That is, we assume that the counterfactuals of the outcome $Y$ are independent of the pair of counterfactuals $\left(M(t'),W(t'')\right)$, and that the treatement $T$ is randomized for the vector of mediators $Z(t)$. Note that these five assumptions can be reduced to the two following assumptions, that we called \textbf{Sequential Ignorability for Multiple Mediators Assumption (SIMMA)}:

$$\begin{array}{rrll}
\eqref{ISB1}  \quad  & \{Y(t,m,w),M(t'),W(t'')\} & \independent & T|X=x, \\
\eqref{ISB4} \quad & Y(t,m,w) & \independent & \left(M(t'),W(t'')\right)|T,X=x.
\end{array}$$

 As a matter of fact we have the following implications:

\begin{eqnarray*}
\eqref{ISB4} & \Rightarrow & Y(t',m,w) \independent  (M(t),W(t)) | T, X=x, \, \forall m,w,t,t' \\
& \Rightarrow  & \left\{\begin{array}{ccl} Y(t',m,w)  & \independent & M(t) | T,X=x, \,\forall m,w,t,t' \\
Y(t',m,w)  & \independent & W(t) | T,X=x, \,\forall m,w,t,t'
\end{array} \right. \\
& \Rightarrow  & \left\{\begin{array}{ccl} Y(t',m,W(t'))  & \independent & M(t) | T,X=x, \,\forall m,t,t' \\
Y(t',M(t'),w)  & \independent & W(t) | T,X=x, \,\forall w,t,t'
\end{array} \right. \,  \footnotemark\\
\eqref{ISB4} & \Rightarrow & \left\{\begin{array}{ccl} \eqref{ISB2}  \\
\eqref{ISB3} 
\end{array} \right.
\end{eqnarray*}
\addtocounter{footnote}{0}
\footnotetext{Independence is all the more true for $w= W(t')$ et $m=M(t')$.}

\begin{eqnarray*}
\eqref{ISB1} & \Rightarrow & \left\{Y(t',m,w),M(t),W(t)\right\} \independent  T| X=x, \\
 & \Rightarrow & \left\{Y(t',m,w),Z(t)\right\} \independent  T| X=x,\\
  & \Rightarrow & Z(t) \independent  T| X=x.\\
  \eqref{ISB1} & \Rightarrow & \eqref{ISB5}.
\end{eqnarray*}

\section{Proof of Theorem 3.1} \label{ptheo1J}

We prove the equations giving the indirect effect of the mediator of interest and the joint indirect effect. The proofs for the direct and total effects can be found in the \href{http://helios.mi.parisdescartes.fr/~ajerolon/indexang.html}{supplementary materials}.

\subsection{Indirect effect via the mediator of interest}
  It follows from the definition that:
\begin{eqnarray}
\delta(t) & = & \mathbb{E}\left[Y(t,M(1),W(t))\right] - \mathbb{E}\left[Y(t,M(0),W(t))\right] \nonumber\\
& = & \int \mathbb{E}\left[Y(t,M(1),W(t))|X=x\right] - \mathbb{E}\left[Y(t,M(0),W(t))|X=x\right] \mathrm{d}F_{X}(x). \nonumber
\end{eqnarray}

  It is then sufficient to demonstrate that:

\begin{eqnarray*}
\mathbb{E}\left[Y(t,M(t'),W(t)|X=x\right]& = & \int_{\mathbb{R}^K} \mathbb{E}\left[Y|M=m,W=w,T=t,X=x\right]\\
&  & \quad \mathrm{d}F_{\left(M(t'),W(t)\right)|X=x}(m,w).
\end{eqnarray*}

 We have:

\begin{equation*}
\begin{array}{rclc}
&  & \mathbb{E}\left[Y\left(t,\left(M(t'),W(t)\right)\right)|X=x\right]  \nonumber\\
                                            & = &\displaystyle\int_{\mathbb{R}^K} \mathbb{E}\left[Y\left(t,m,w\right)|\left(M(t'),W(t)\right)=(m,w),X=x\right] \mathrm{d}F_{\left(M(t'),W(t)\right)|X=x}(m,w)\\
& = &  \displaystyle\int_{\mathbb{R}^K} \mathbb{E}\left[Y\left(t,m,w\right)|\left(M(t'),W(t)\right)=(m,w),T=t,X=x\right]\\
&  & \mathrm{d}F_{\left(M(t'),W(t)\right)|X=x}(m,w) &\footnotemark \\
& = & \displaystyle\int_{\mathbb{R}^K} \mathbb{E}\left[Y(t,m,w)|\left(M(t),W(t)\right)=(m,w),T=t,X=x\right]\\
&   & \mathrm{d}F_{(M(t'),W(t))|X=x}(m,w) & \footnotemark\\
& = & \displaystyle\int_{\mathbb{R}^K} \mathbb{E}\left[Y|M=m,W=w,T=t,X=x\right]\\
&  & \mathrm{d}F_{(M,W)|X=x}(m,w). & \footnotemark 
\end{array}
\end{equation*}
\addtocounter{footnote}{-2}
\footnotetext{By \eqref{ISB1}.}
\stepcounter{footnote}
\footnotetext{By  \eqref{ISB4}.}
\stepcounter{footnote}
\footnotetext{By consistency relation.}

 In the case where $M$ and $W$ are independent, we have: 

\begin{eqnarray*} 
\mathrm{d}F_{\left(M(t'),W(t)\right)|X=x}(m,w) & = &f_{\left(M(t'),W(t)\right)|X=x}(m,w)\mathrm{d}m\mathrm{d}w \\
& = & f_{M(t')|X=x}(m)\mathrm{d}m f_{W(t)|X=x}(w)\mathrm{d}w\\
& = & f_{M|T=t',X=x}(m)\mathrm{d}m f_{W|T=t,X=x}(w)\mathrm{d}w
\end{eqnarray*}

 We therefore have: 
\begin{equation*}
\begin{array}{clll}
  \delta(t) & = & \displaystyle\int \int_{\mathbb{R}^K} \mathbb{E}\left[Y|M=m,W=w,T=t,X=x\right] \\
  &  & \{f_{M|T=1,X=x}(m)\mathrm{d}m  f_{W|T=t,X=x}(w)\mathrm{d}w-f_{M|T=0,X=x}(m)\mathrm{d}m  f_{W|T=t,X=x}(w)\mathrm{d}w\}\\
  &  & \mathrm{d}F_{X}(x)\\
  & = & \displaystyle\int \int \int_{\mathbb{R}^{K-1}} \mathbb{E}\left[Y|M=m,W=w,T=t,X=x\right] \\
  &  & f_{W|T=t,X=x}(w)\mathrm{d}w\{f_{M|T=1,X=x}(m) -f_{M|T=0,X=x}(m) \}\mathrm{d}m \mathrm{d}F_{X}(x)\\
  & = & \displaystyle\int \int \mathbb{E}\left[Y|M=m,T=t,X=x\right] \{f_{M|T=1,X=x}(m) -f_{M|T=0,X=x}(m) \}\mathrm{d}m \mathrm{d}F_{X}(x) & \footnotemark\\
   & = & \displaystyle\int \int \mathbb{E}\left[Y|M=m,T=t,X=x\right] \{\mathrm{d} F_{M|T=1,X=x}(m) -\mathrm{d} F_{M|T=0,X=x}(m) \} \mathrm{d}F_{X}(x).\\
  \end{array}
  \end{equation*}
\addtocounter{footnote}{0}
\footnotetext{By the formula of total expectations $\mathbb{E}\left[X\right]=\mathbb{E}\left[\mathbb{E}\left[X|A\right]\right]=\sum^n_{i=1}\mathbb{E}\left[X|A\right]\mathbb{P}\left(A\right)$, where $A$ is a partition.}

\subsection{Joint indirect effect}

It follows from the definition that
\begin{eqnarray}
\delta^Z(t) & = & \mathbb{E}\left[Y(t,Z(1))\right] - \mathbb{E}\left[Y(t,Z(0))\right] \nonumber\\
& = & \int \mathbb{E}\left[Y(t,Z(1))|X=x\right] - \mathbb{E}\left[Y(t,Z(0))|X=x\right] \mathrm{d}F_{X}(x) \nonumber
\end{eqnarray}

  It is then sufficient to demonstrate that:

\begin{equation*}
\mathbb{E}\left[Y(t,Z(t'))|X=x\right]= \int_{\mathbb{R}^K} \mathbb{E}\left[Y|Z=z,T=t,X=x\right]\mathrm{d}F_{Z|T=t',X=x}(z).
\end{equation*}
\vspace{0.4cm}
  Previously we showed that for $t',t$ $\in$ $\{0,1\}$:

\begin{eqnarray*}\mathbb{E}\left[Y(t,M(t'),W(t))|X=x\right] & = & \int_{\mathbb{R}^K} \mathbb{E}\left[Y|M=m,W=w,T=t,X=x\right]\\
 &  & \quad \mathrm{d}F_{\left(M(t'),W(t)\right)|X=x}(m,w).
 \end{eqnarray*}

  Taking $t=t'$, this equation becomes:

 \begin{eqnarray*}
 \mathbb{E}\left[Y(t,M(t'),W(t'))|X=x\right]& = & \int_{\mathbb{R}^K} \mathbb{E}\left[Y|M=m,W=w,T=t,X=x\right]\\
 &  & \mathrm{d}F_{\left(M(t'),W(t')\right)|X=x}(m,w) \\
\end{eqnarray*}

That is:

\begin{eqnarray*}
  \mathbb{E}\left[Y(t,Z(t'))|X=x\right]& = & \int_{\mathbb{R}^K} \mathbb{E}\left[Y|Z=z,T=t,X=x\right]\\
 &  & \mathrm{d}F_{\left(Z(t')\right)|X=x}(m,w) \\
& = & \int_{\mathbb{R}^K} \mathbb{E}\left[Y|Z=z,T=t,X=x\right]\mathrm{d}F_{Z|T=t',X=x}(z). \\
\end{eqnarray*}


\section{Models}\label{model}
 We give here the models used for the simulation study in Subsection 4.\ref{ss:empirical_study}.
\subsection*{Model 1: Continuous outcome and continuous mediators}
\begin{itemize}

\item $T$ follows a Bernoulli distribution $\mathcal{B}(0.3)$.
\item The joint distribution of the two counterfactual mediators is
\begin{eqnarray*}
\left(\begin{array}{c}
M^1(t)   \\
M^2(t)  \\
\end{array}\right) & \sim &  \mathcal{N}\left(\mu=\left(\begin{array}{c}
1 +4t \\
2 +6 t\\
\end{array}\right),\Sigma\right).
\end{eqnarray*}
\item The counterfactual outcome follows the normal distribution
\begin{eqnarray*}
Y\left(t,M^1(t'),M^2(t'')\right)  & \sim &  \mathcal{N}(1+10  t + 5  M^1(t') + 4  M^2(t'') ,1).
\end{eqnarray*}
\end{itemize}

\medskip
In table \ref{truevalue1}, we show the real values of causal effects entailed by model 1.

\begin{table}[htp]
\centering
\begin{tabular}{|c|c|c|c|c|}
\hline
$\delta^Z$ & $\delta^1$  & $\delta^2$ & $\zeta$  & $\tau$ \\
   \hline
 44 & 20  & 24 & 10 & 54 \\ 
   \hline
\end{tabular}
\caption{Real values of the causal effects entailed by model 1.}\label{truevalue1}
\end{table}


\subsection*{Model 2: Binary outcome (logit) with continuous mediators}
\begin{figure}
\vspace{-3cm}
\centering
\includegraphics[height=0.3\textwidth]{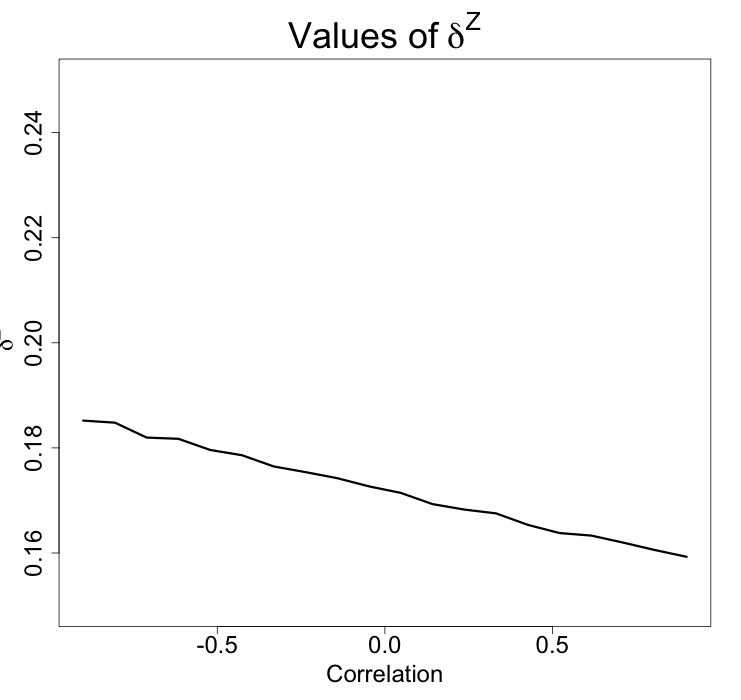}
\includegraphics[height=0.3\textwidth]{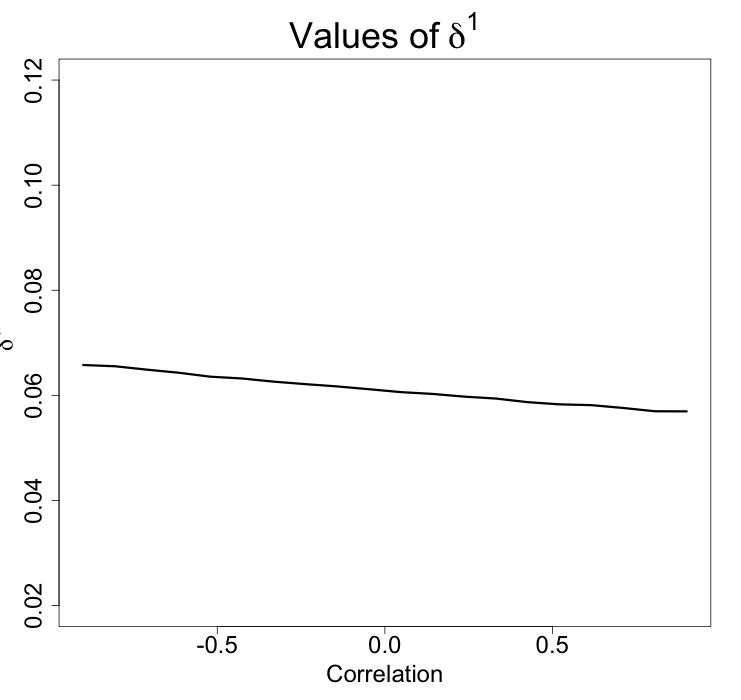}
\includegraphics[height=0.3\textwidth]{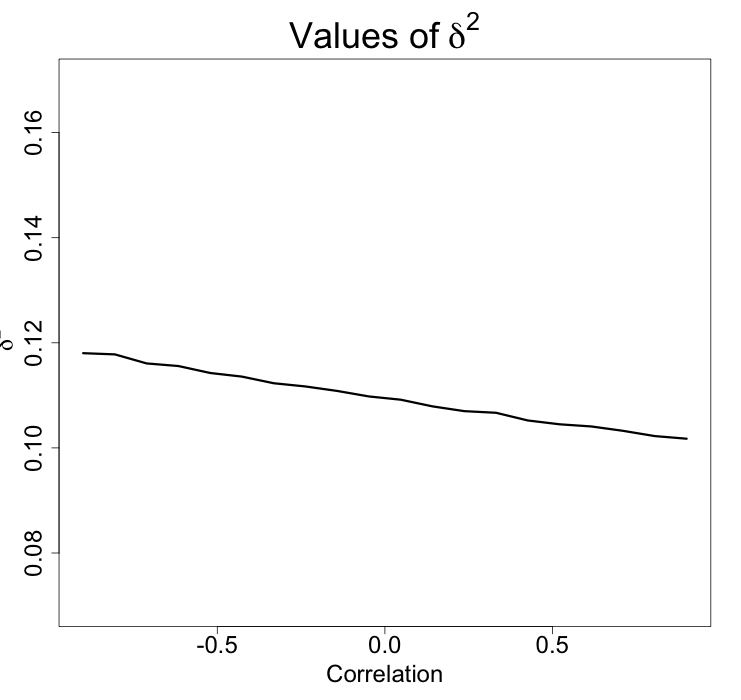}
\includegraphics[height=0.3\textwidth]{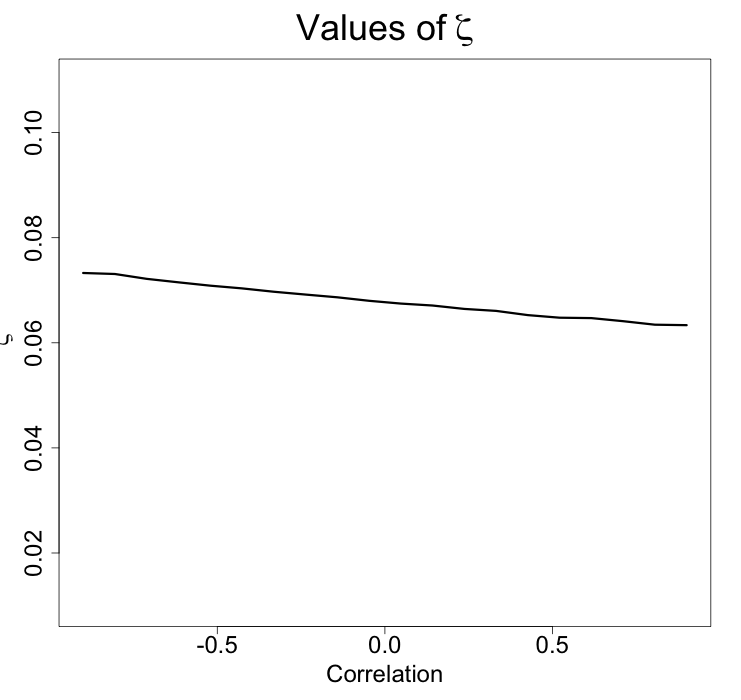}
\includegraphics[height=0.3\textwidth]{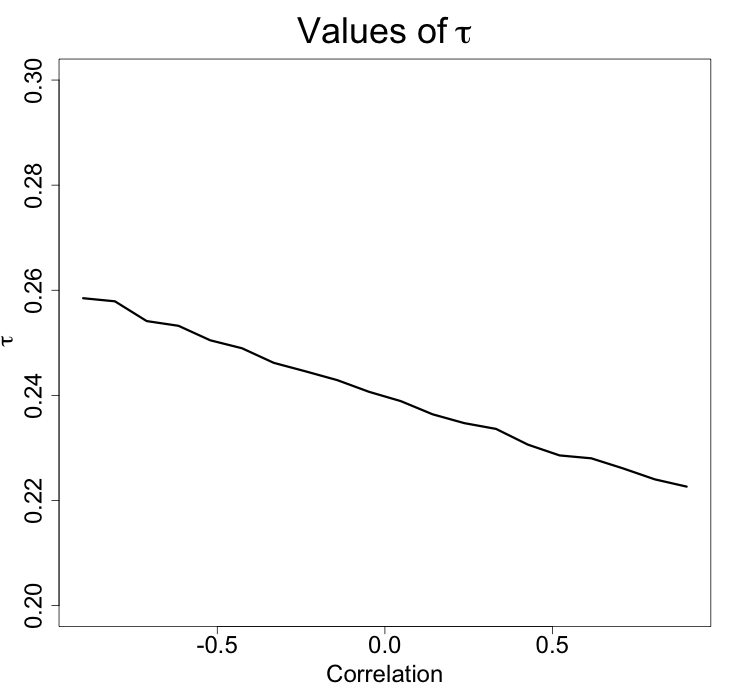}
\caption{Variation in causal effects, with binary outcome, due to correlation}\label{truevalue2}
\end{figure}

\begin{itemize}
\item Treatment: $T\sim\mathcal{B}(0.3)$.
\item Counterfactuals mediators:
\begin{eqnarray*}
\left(\begin{array}{c}
M^1(t)   \\
M^2(t)  \\
\end{array}\right) & \sim &  \mathcal{N}\left(\mu=\left(\begin{array}{c}
 0.1+ 0.6  t  \\
 0.2+ 0.8  t  
\end{array}\right),\Sigma\right).
\end{eqnarray*}
\item The counterfactual outcome follows the logistic regression:
\begin{eqnarray*}
Y(t,M^1(t'),M^2(t'')) & \sim & B\left(\cfrac{1}{1+ \exp\left(-2 + 0.4  t +0.6M^1(t')+ 0.8 M^2(t'')\right)}\right).
\end{eqnarray*}
\end{itemize}
With this choice of parameters, 30\% of the sampled observations are cases. As we can see in Corollary \ref{cor2J}, with binary outcome, causal effects are related to the covariance of mediators. Figure \ref{truevalue2} shows how the true causal values change when correlation changes.
 
\section{Complementary results}\label{result}

\begin{figure}[htp]
\centering
\includegraphics[height=0.5\textwidth]{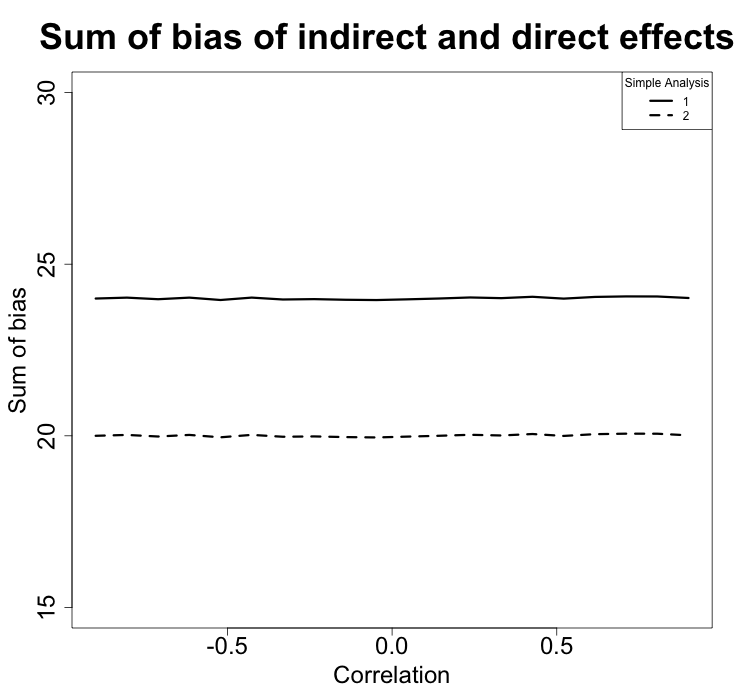}
\caption{Adding the bias for the direct and indirect effects presented in Figure \ref{resultsim}. Estimates were obtained by simple analyses on data simulated according to model 1. See table \ref{truevalue1} for the true values of the effects.}\label{sumbias}
\end{figure}

\begin{landscape}
\begin{figure}
\vspace{-3cm}
\centering
\includegraphics[height=0.3\textwidth]{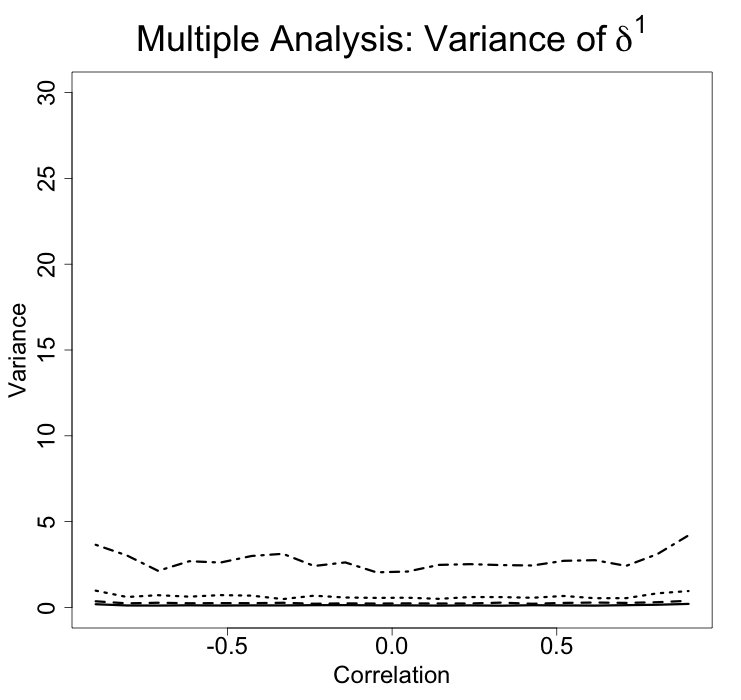}
\includegraphics[height=0.3\textwidth]{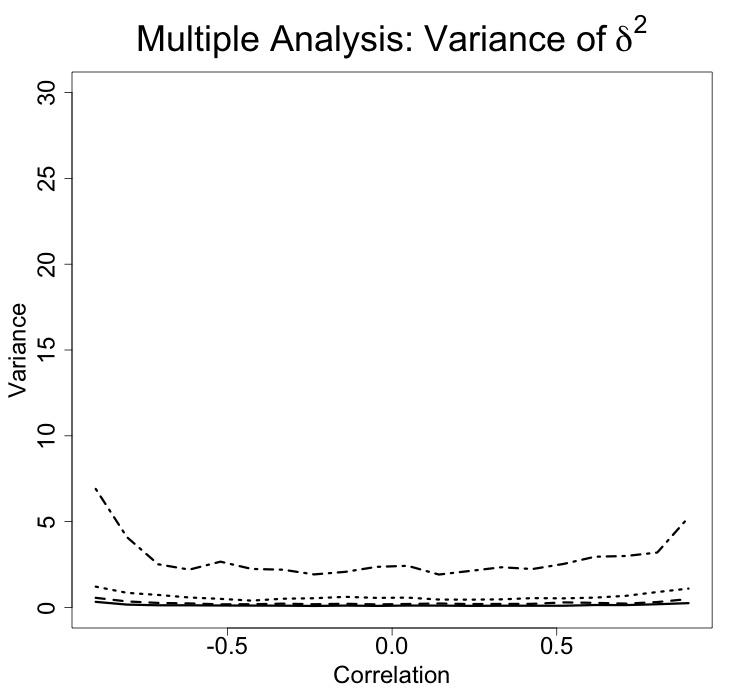}
\includegraphics[height=0.3\textwidth]{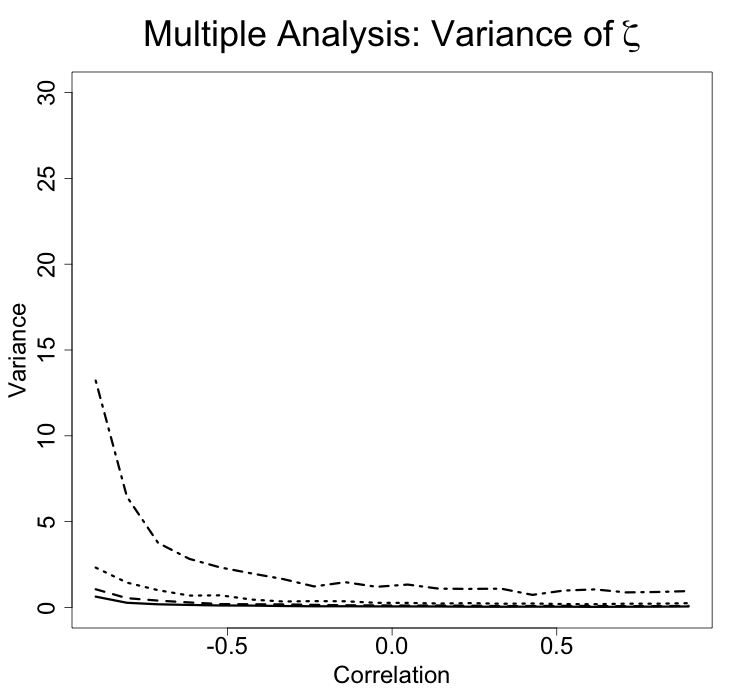}
\includegraphics[height=0.3\textwidth]{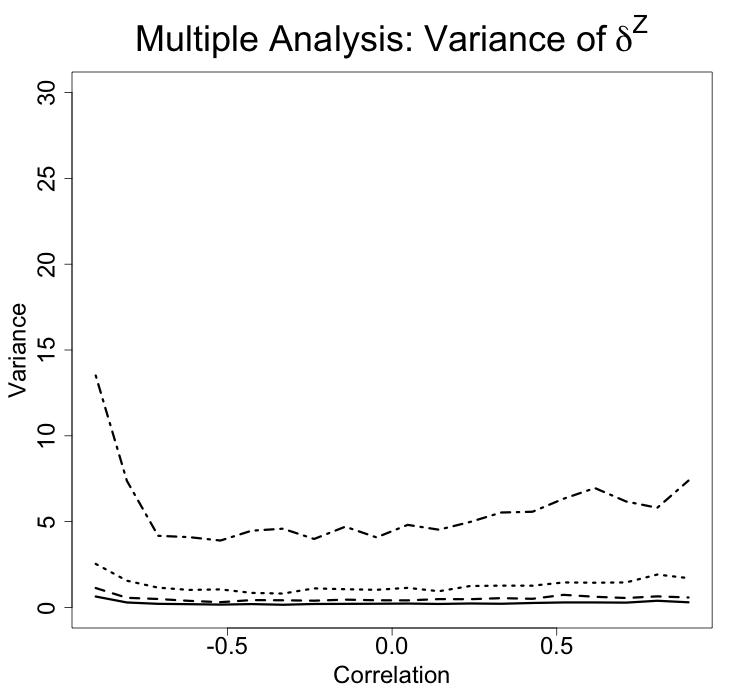}
\includegraphics[height=0.3\textwidth]{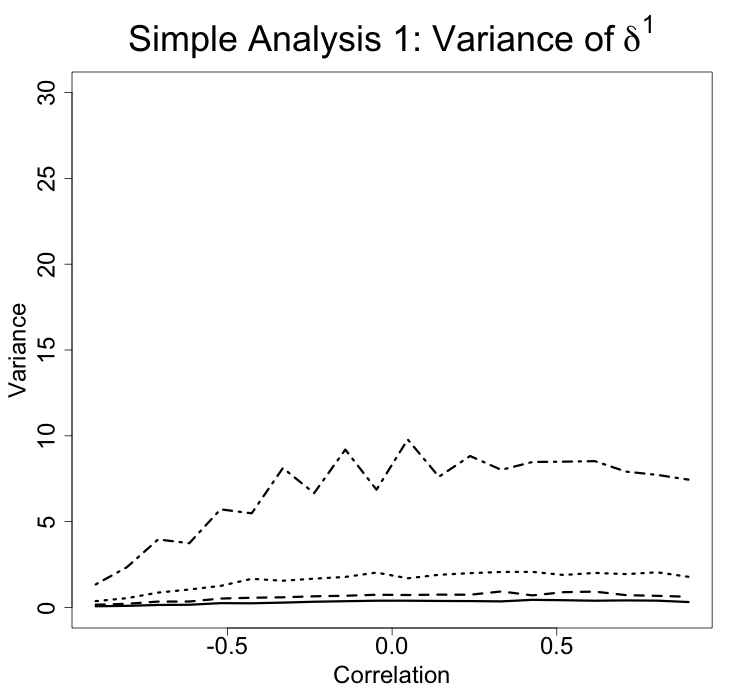}
\includegraphics[height=0.3\textwidth]{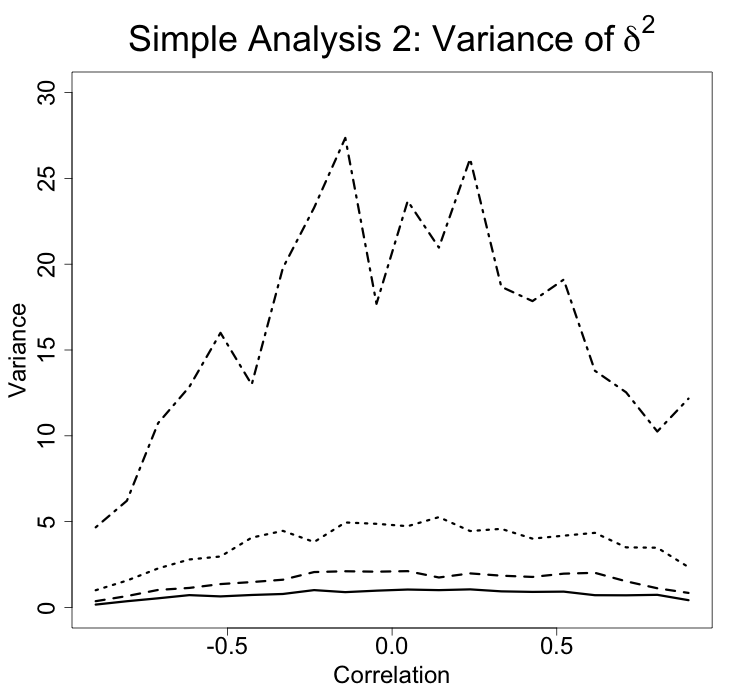}
\includegraphics[height=0.3\textwidth]{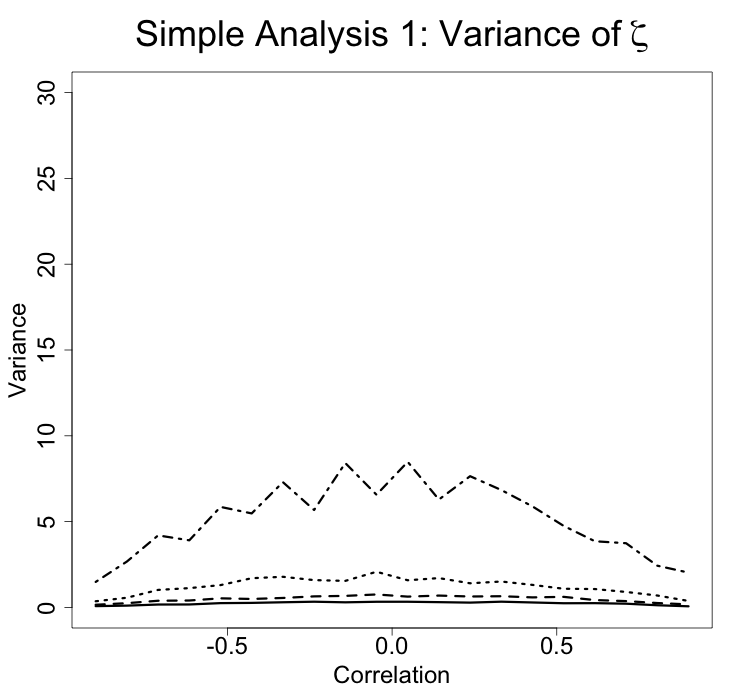}
\includegraphics[height=0.3\textwidth]{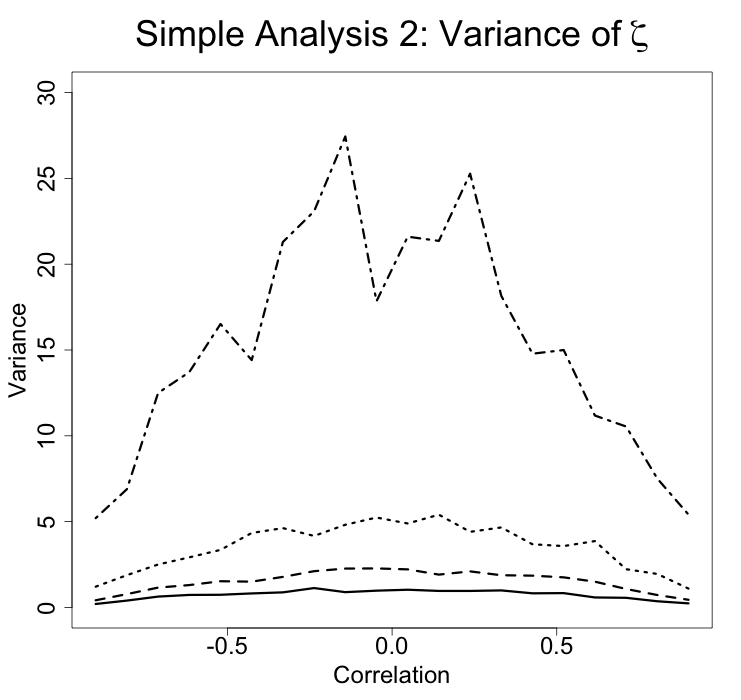}
\includegraphics[height=0.3\textwidth]{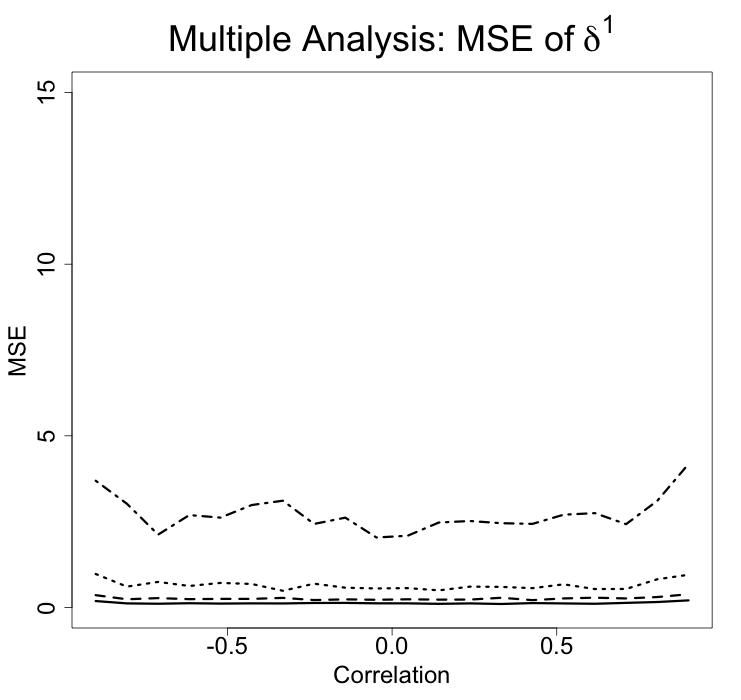}
\includegraphics[height=0.3\textwidth]{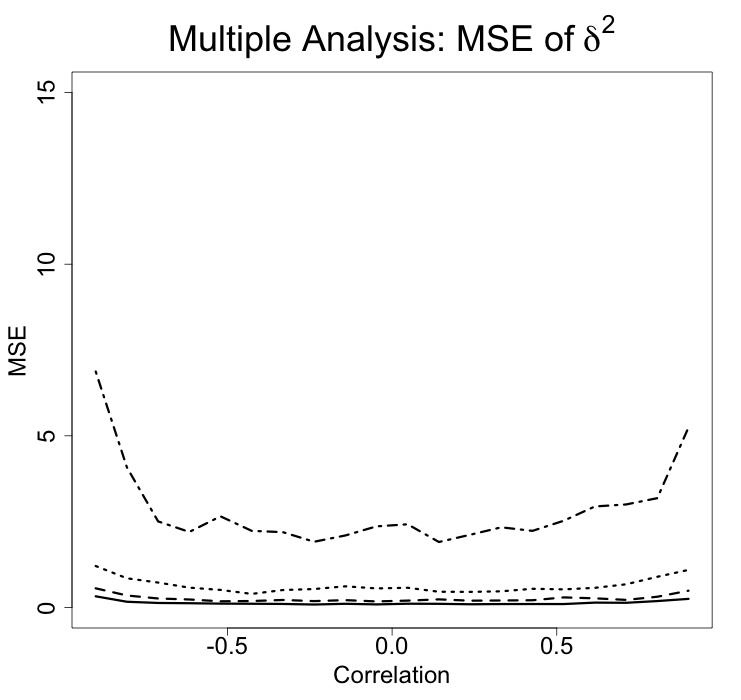}
\includegraphics[height=0.3\textwidth]{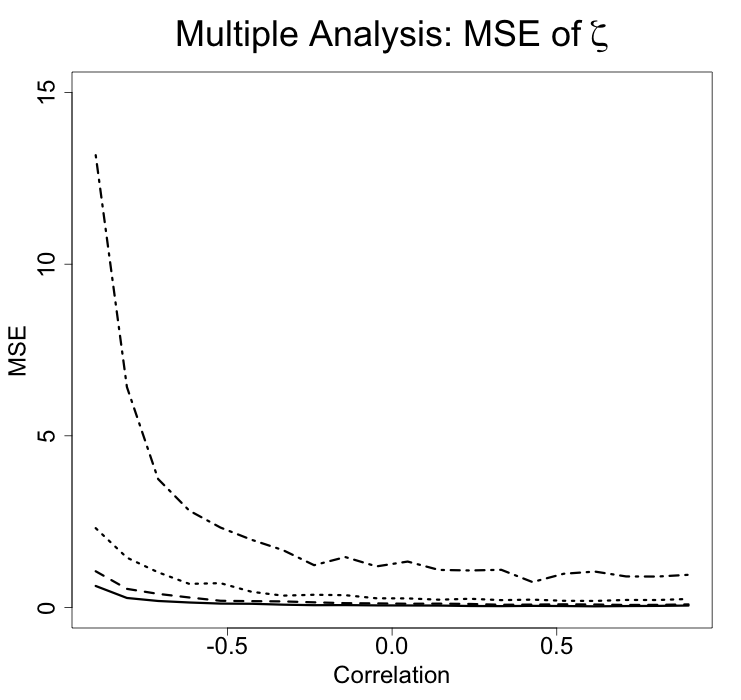}
\includegraphics[height=0.3\textwidth]{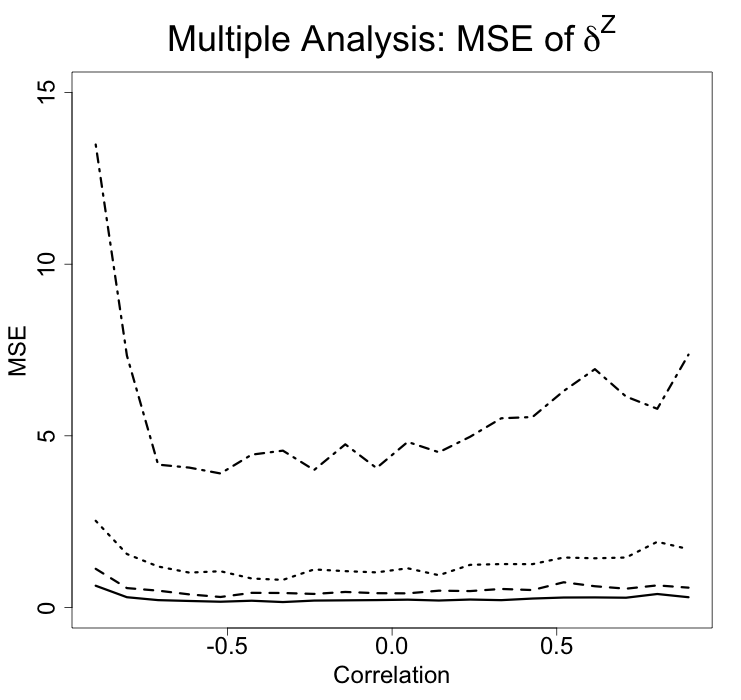}
\includegraphics[height=0.3\textwidth]{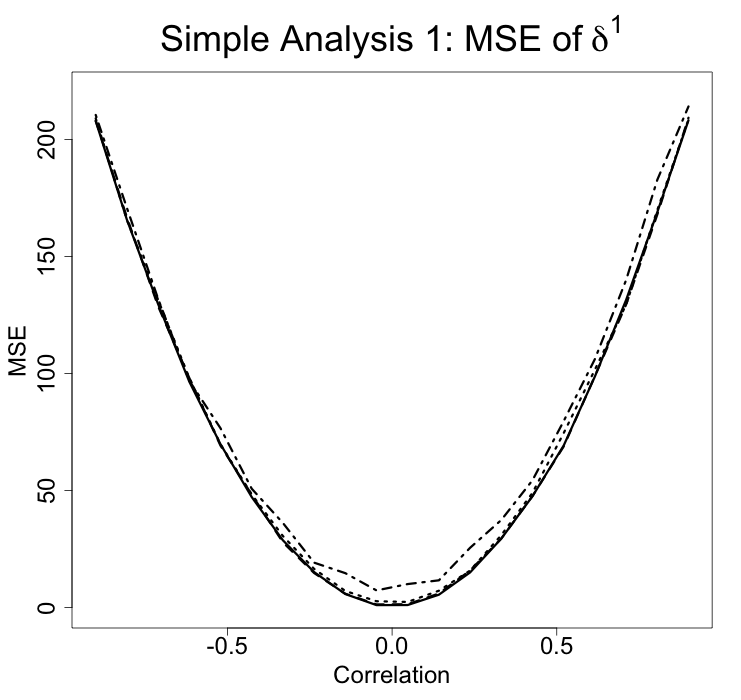}
\includegraphics[height=0.3\textwidth]{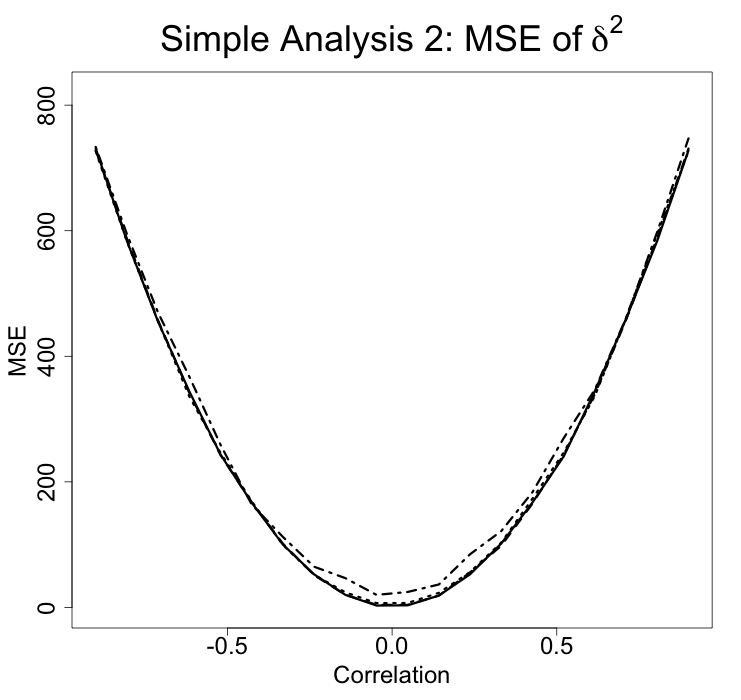}
\includegraphics[height=0.3\textwidth]{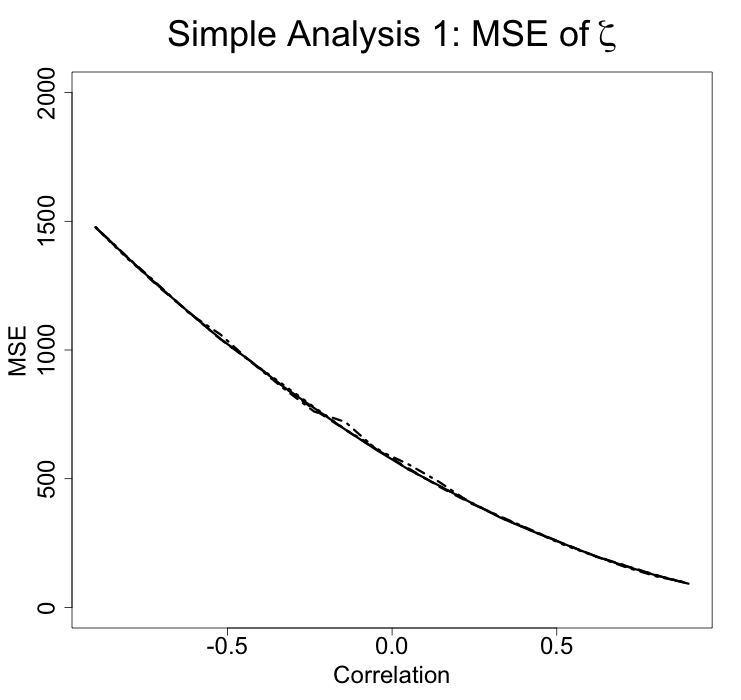}
\includegraphics[height=0.3\textwidth]{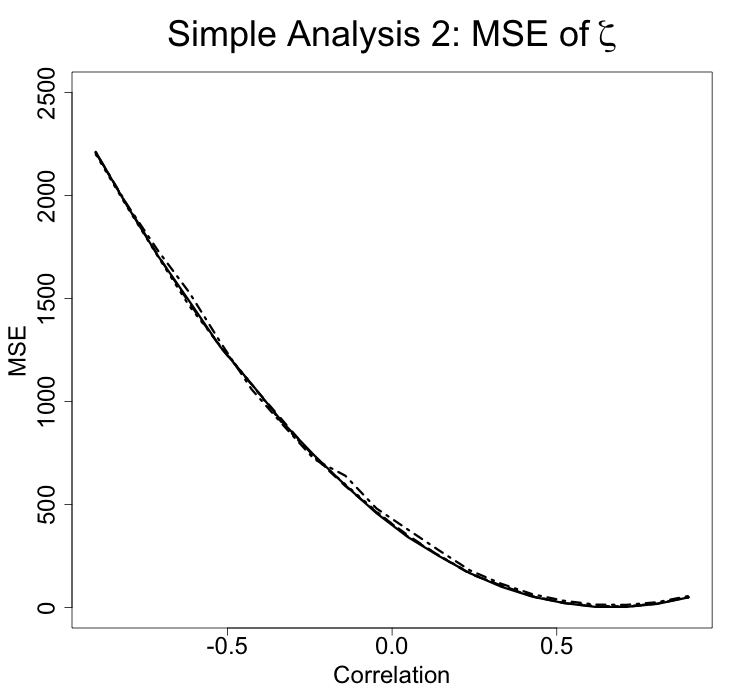}
\includegraphics[height=0.05\textwidth]{legend.png}
\caption{Variance and MSE of indirect and direct effect estimators when the correlation between mediators varies in model 1. The first two rows (multiple analysis in row 1 and simples analyses in row 2) represents the variance and the last two rows the MSE (multiple analysis in row 3 and simples analyses in row 4).}\label{resultsimother}
\end{figure}
\end{landscape}

\begin{landscape}
\begin{figure}
\vspace{-3cm}
\centering
\includegraphics[height=0.3\textwidth]{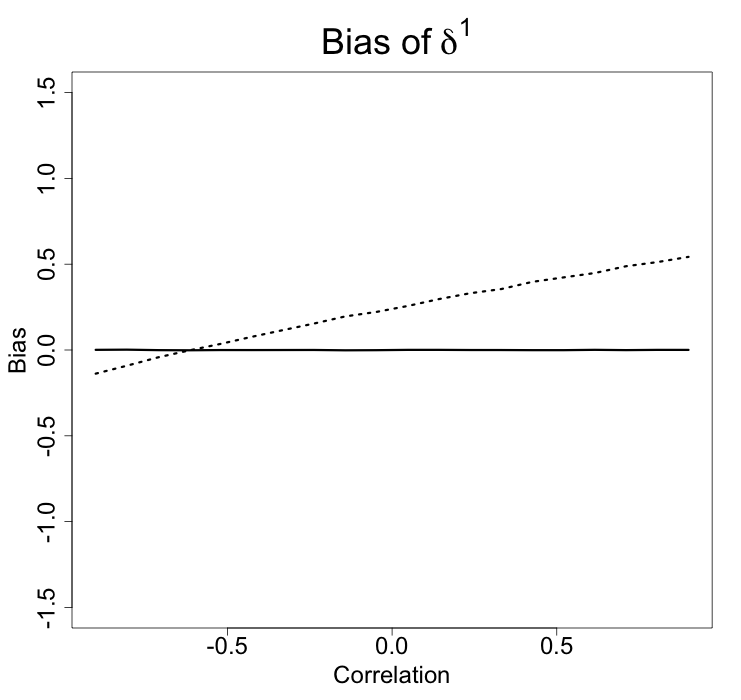}
\includegraphics[height=0.3\textwidth]{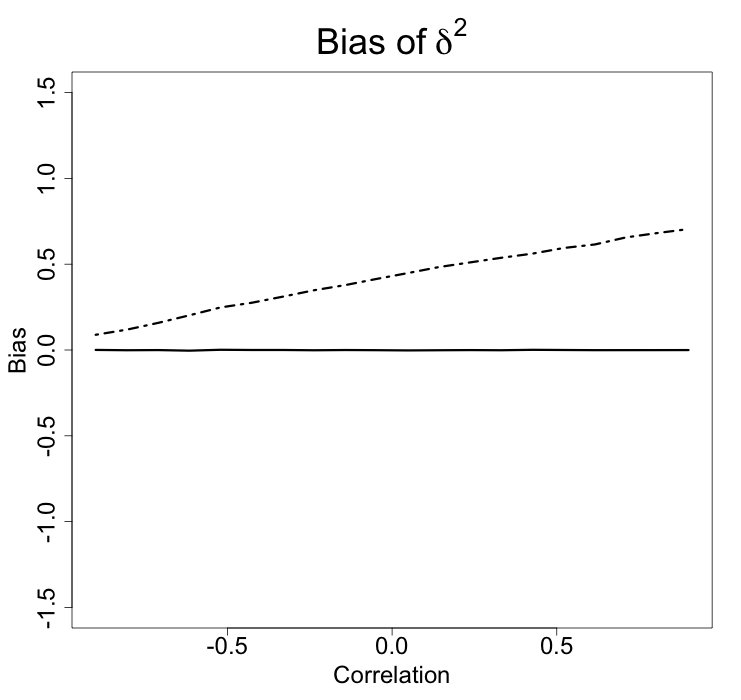}
\includegraphics[height=0.3\textwidth]{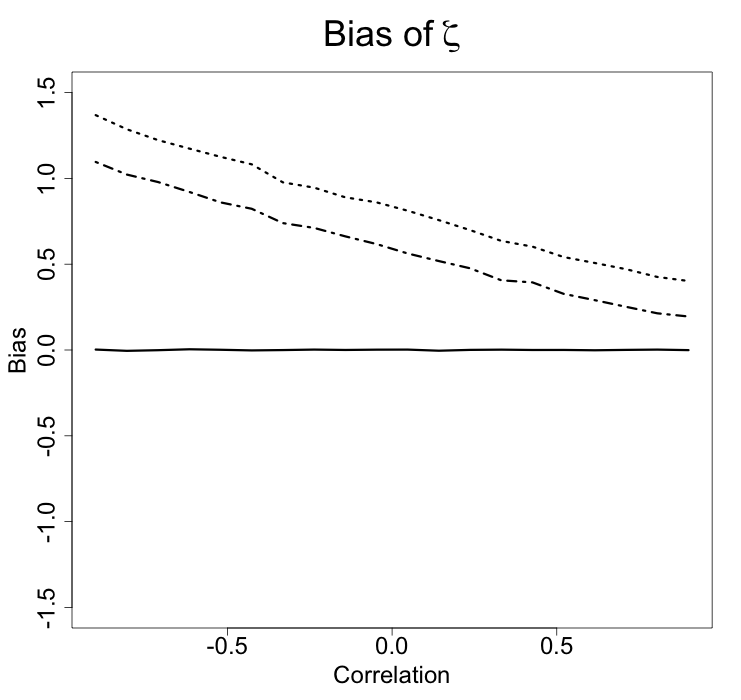}
\includegraphics[height=0.3\textwidth]{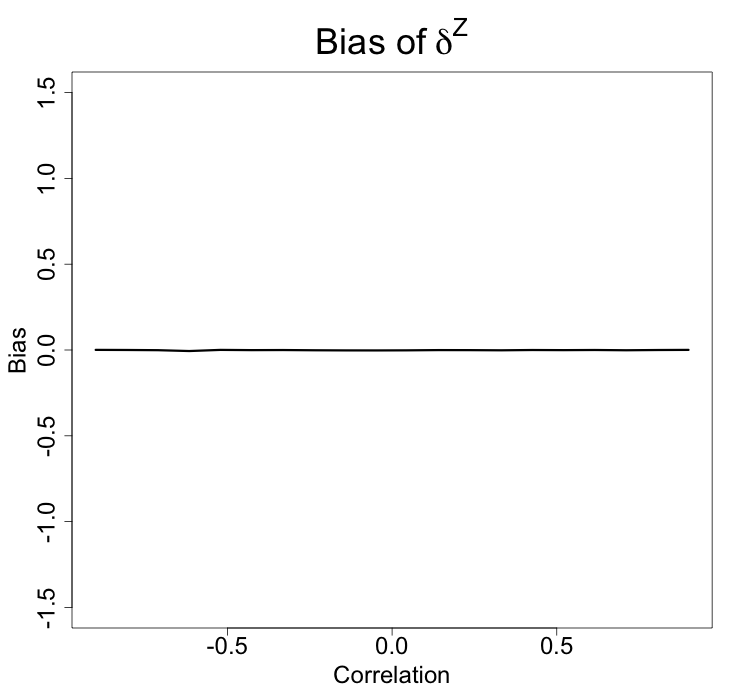}
\includegraphics[height=0.3\textwidth]{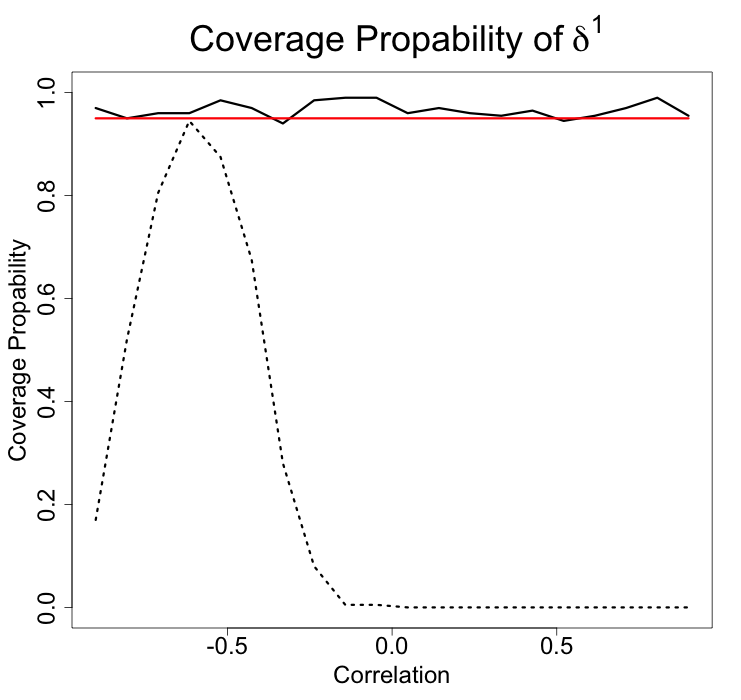}
\includegraphics[height=0.3\textwidth]{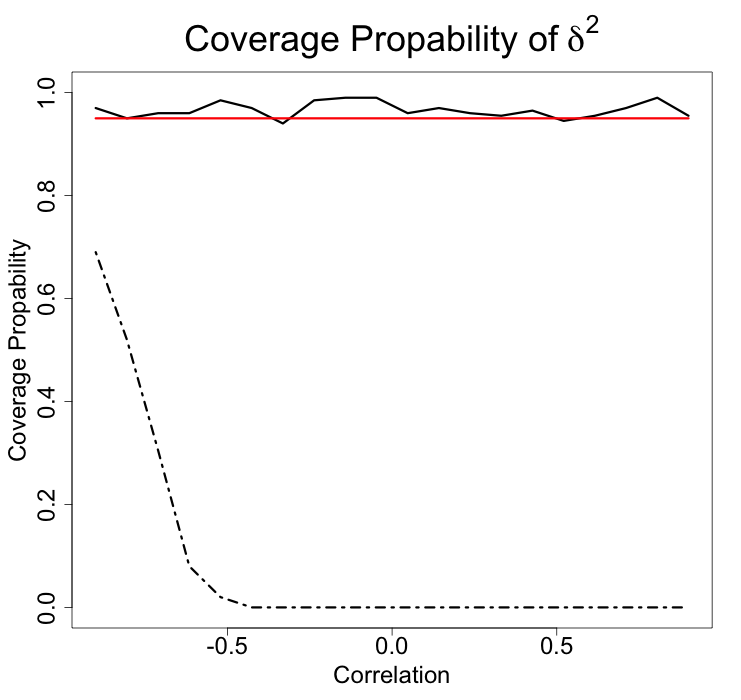}
\includegraphics[height=0.3\textwidth]{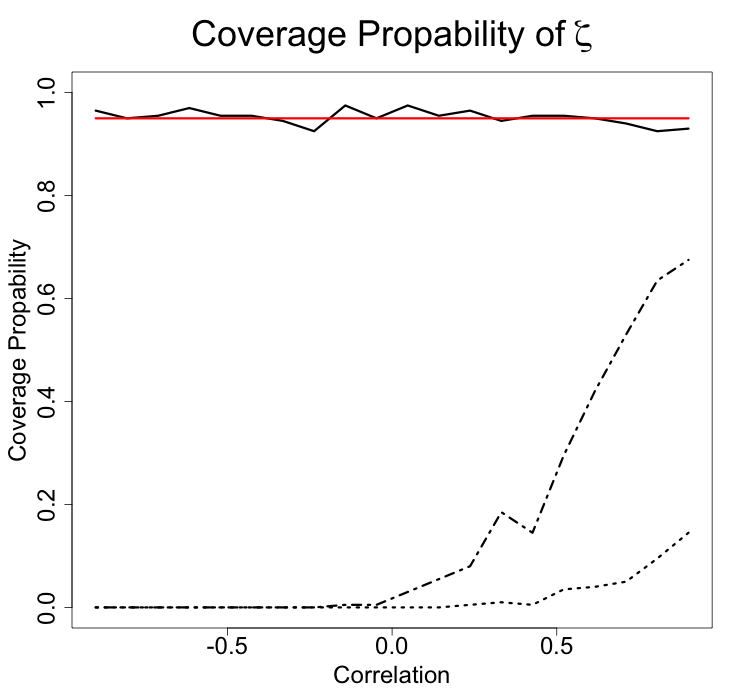}
\includegraphics[height=0.3\textwidth]{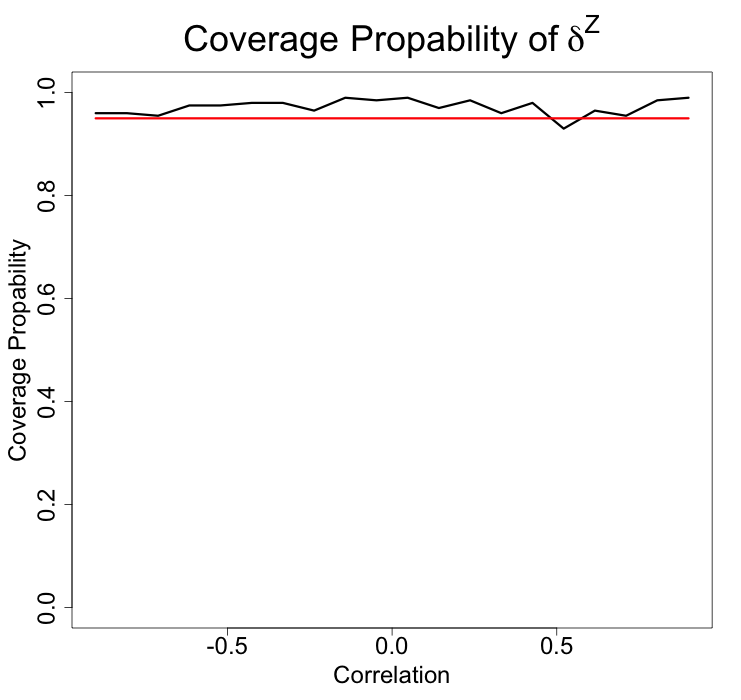}
\includegraphics[height=0.3\textwidth]{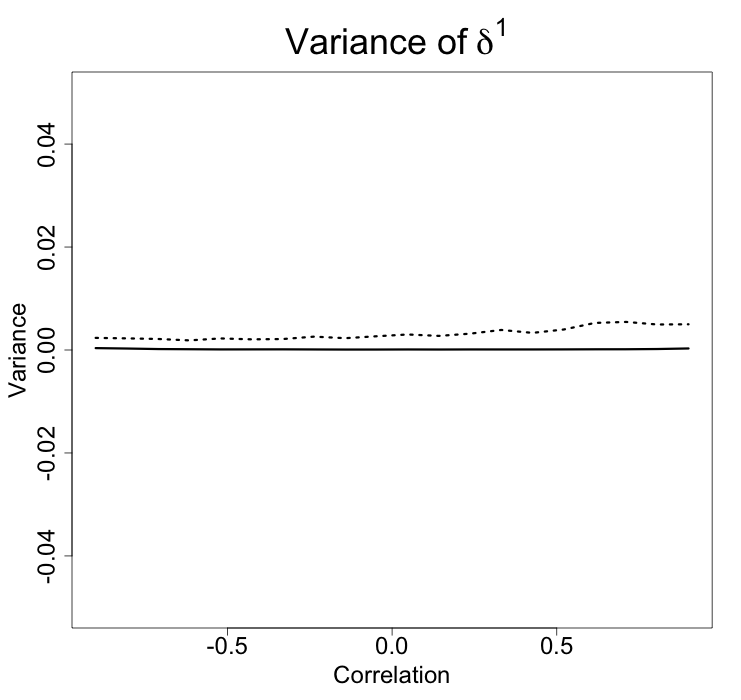}
\includegraphics[height=0.3\textwidth]{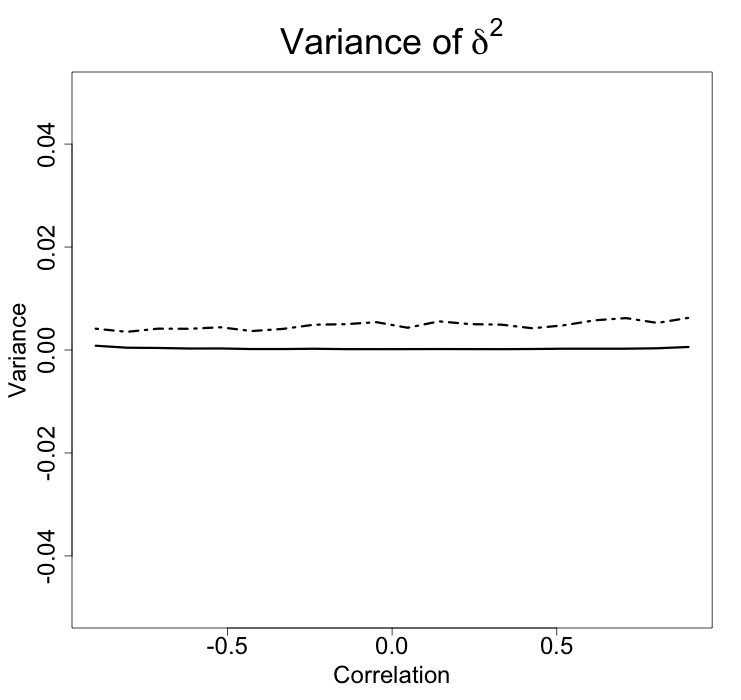}
\includegraphics[height=0.3\textwidth]{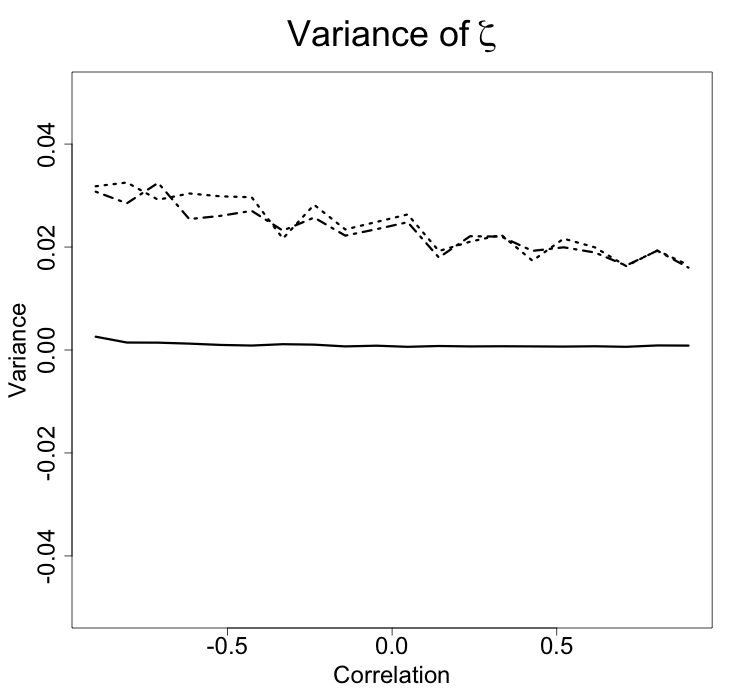}
\includegraphics[height=0.3\textwidth]{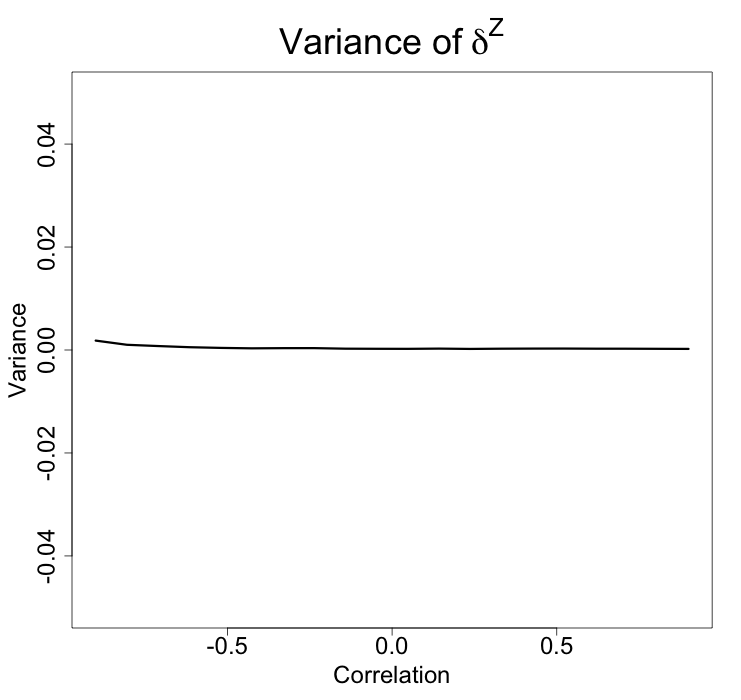}
\includegraphics[height=0.3\textwidth]{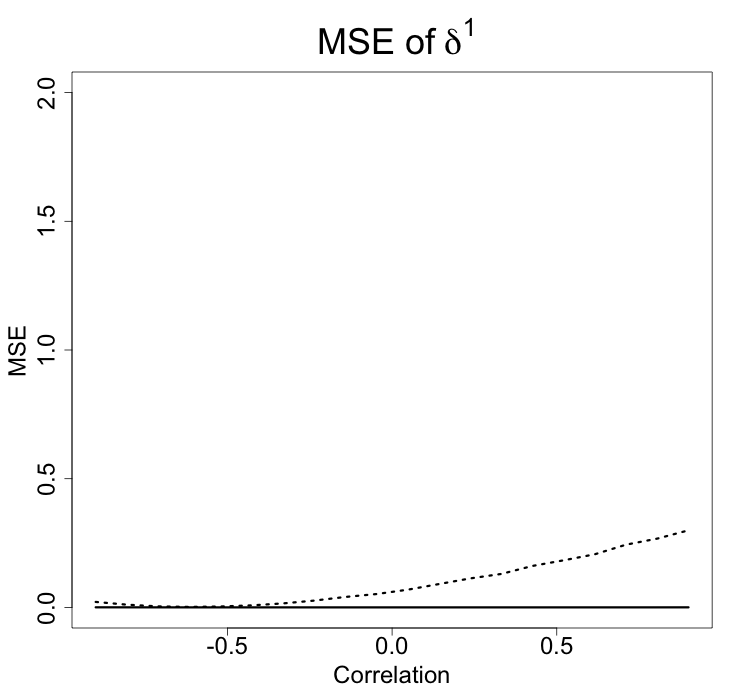}
\includegraphics[height=0.3\textwidth]{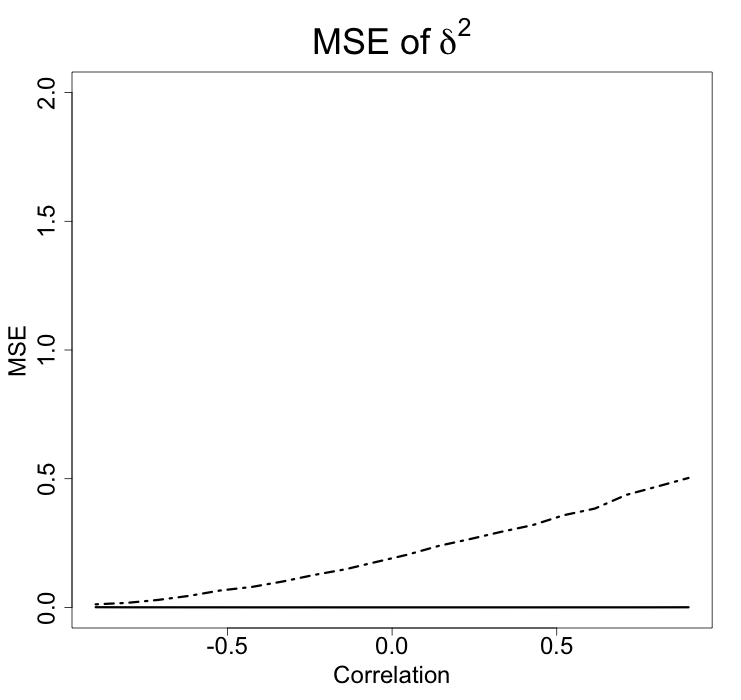}
\includegraphics[height=0.3\textwidth]{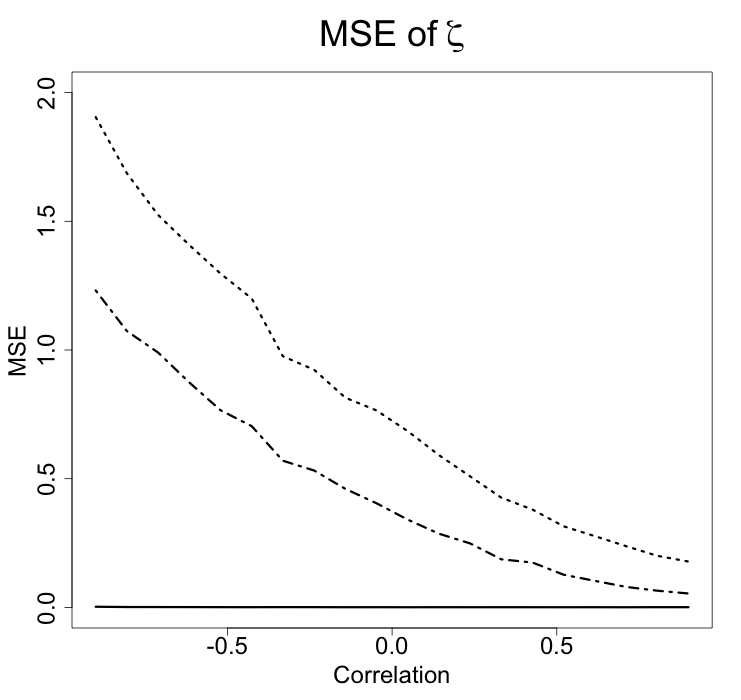}
\includegraphics[height=0.3\textwidth]{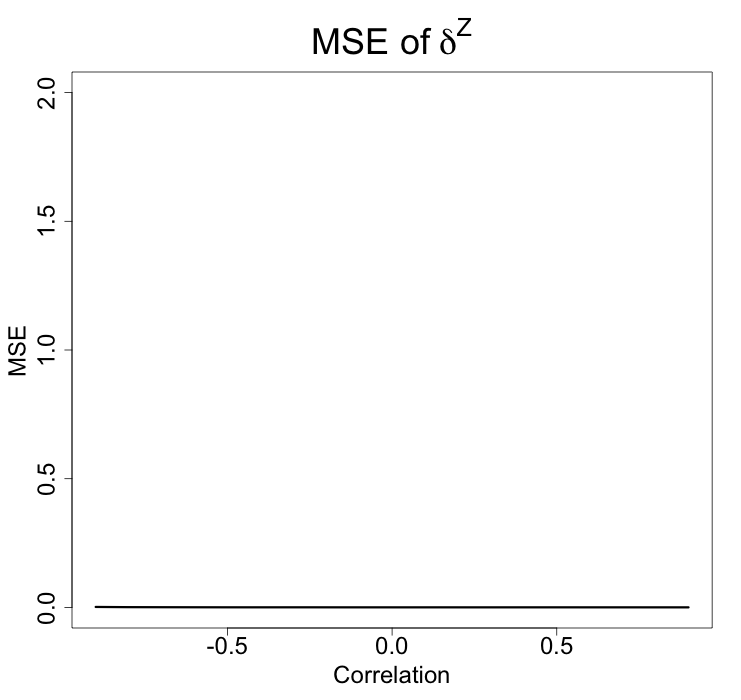}
\includegraphics[height=0.05\textwidth]{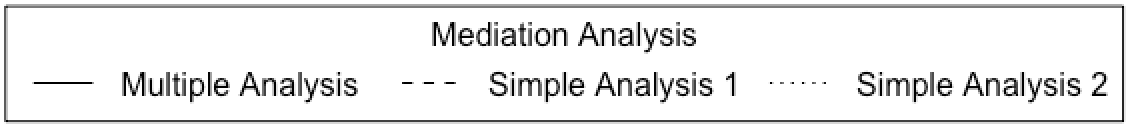}
\caption{Bias (first row), Coverage probability (second row), Variance (third row) and MSE (fourth row) of the indirect and direct effect estimators when the correlation between mediators varies in model 2.}\label{resultsimeff}
\end{figure}
\end{landscape}

\bibliographystyle{DeGruyter}
\bibliography{Biblio}
\end{document}